\documentclass[10pt,prb,aps,twocolumn,floatfix]{revtex4}

\usepackage{graphicx,float}
\usepackage{amssymb}
\usepackage{amsmath}
\usepackage{subfigure}
\usepackage [usenames,dvipsnames] {color}
\usepackage{amsmath,hyperref}

\usepackage{amsbsy}
\usepackage{amsmath}
\usepackage{epsfig}
\newcommand\be{\begin{eqnarray}}
\newcommand\ee{\end{eqnarray}}
\newcommand\beq{\begin{equation}}
\newcommand\eeq{\end{equation}}

\begin{document}

\title{Periodically driven integrable systems with long-range pair potentials}

\author{Sourav Nandy, K. Sengupta, and Arnab Sen}

\affiliation{Department of Theoretical Physics, Indian Association
for the Cultivation of Science, Jadavpur, Kolkata 700032, India.}

\begin{abstract}

We study periodically driven closed systems with a long-ranged
Hamiltonian by considering a generalized Kitaev chain with pairing
terms which decay with distance as a power law characterized by
exponent $\alpha$. Starting from an initial unentangled state, we
show that all local quantities relax to well-defined steady state
values in the thermodynamic limit and after $n \gg 1$ drive cycles
for any $\alpha$ and driving frequency $\omega$. We introduce a
distance measure, $\mathcal{D}_l(n)$, that characterizes the
approach of the reduced density matrix of a subsystem of $l$ sites
to the reduced density matrix of the final steady state. We chart
out the $n$ dependence of ${\mathcal D}_l(n)$ and identify a
critical value $\alpha=\alpha_c$ (which depends only on the
time-averaged Hamiltonian) below which they generically decay to
zero as $(\omega/n)^{1/2}$. For $\alpha > \alpha_c$, in contrast,
${\mathcal D}_l(n) \sim (\omega/n)^{3/2}[(\omega/n)^{1/2}]$ for
$\omega \to \infty [0]$ with at least one intermediate dynamical
transition. An identical behavior is found for relaxation of all
non-trivial correlation functions of the model to their steady-state
values. We also study the mutual information propagation to
understand the nature of the entanglement spreading in space with
increasing $n$ for such long-ranged systems. We point out existence
of qualitatively new features in the space-time dependence of mutual
information for $\omega <  \omega^{(1)}_c$, where $\omega^{(1)}_c$
is the largest critical frequency for the dynamical transition for a
given $\alpha$. One such feature is the presence of {\it multiple}
light cone-like structures which persists even when $\alpha$ is large.
We also show that the nature of space-time dependence of the
mutual information of long-ranged Hamiltonians with $\alpha < 2$
differs qualitatively from their {\it{short-ranged counterparts}}
with  $\alpha > 2$ for any drive frequency and relate this
difference to the behavior of the Floquet group velocity of such
driven system.

\end{abstract}

\date{\today}

\maketitle

\section{Introduction and motivation}
\label{intro}

Recent experimental progress in manipulating well-isolated quantum
systems such as ultracold quantum gases~\cite{Blochreview2005,
Goldman2014, Langenetalreview2015, Eck17} and trapped ion
systems~\cite{LeibfriedBMW2003, MoehringMOYMDM2007, KimCKIEFLDM2010,
DuanM2010} has led to a renewed interest in closed many-body systems
driven by purely unitary dynamics. Even though the system is not
connected to any external heat bath and is thus always in a pure
quantum state, it has now been understood that the increase and
spreading of quantum
entanglement~\cite{PopescuSW2006,NandkishoreH2015} between its
degrees of freedom as a function of time due to the time-dependence
of some parameter of the system's Hamiltonian leads to
the necessity of a mixed density matrix description for any
subsystem. This, in turn, leads to the possibility of well-defined
steady states at late times~\cite{CalabreseC2006, KollathLA2007,
RigolDO2008, MoeckelK2008, PolkovnikovSSV2011, LazaridesAM2014a,
LazaridesAM2014b, AlessioR2014, PonteCPA2015} for the subsystem's
properties as long as the rest of the system (which we call
``environment'' henceforth) is much bigger. Thus, the nature of
entanglement propagation in these far-from-equilibrium regimes is
central to their complete understanding. Systems that are
continually driven by a periodic drive in time are of particular
interest since these are known to lead to non-equilibrium states
that have no equilibrium counterparts, e.g., Floquet time crystals
in many-body localized systems~\cite{ElseBN2016, KhemaniLMS2016} and
dynamical topological ordering~\cite{KitagawaBRD2010, NathanR2015}.

The propagation of quantum entanglement in non-relativistic systems
with short-ranged interactions is a well-studied subject by now. The
seminal work of Lieb and Robinson~\cite{LiebR1972} showed the
existence of a maximum velocity of propagation for correlations in
translationally invariant spin systems with nearest neighbor
interactions which also places a bound on the rate of entanglement
propagation. In integrable systems, entanglement propagates
ballistically~\cite{CalabreseC2005} when the quantum dynamics is
started from an initial unentangled state and the resulting ``light
cone effect'' (see Ref.~\onlinecite{CheneauBPESFGBKK2012} for
experimental observation of this effect) is caused by the
propagation of entangled quasiparticle pairs at finite velocities.
Recent studies have now demonstrated that this ballistic spreading
of entanglement may be more generic and is also present in
non-integrable systems~\cite{KimH2013}. Global quantum quenches,
where some parameter of the Hamiltonian is instantaneously changed
to another value and the state is then propagated with the new
Hamiltonian, provide possibly the simplest setup to study such
entanglement propagation.

Less is known about entanglement propagation in long-ranged systems
where it is expected that qualitatively different features should
arise due to the non-locality of the interactions. The first
generalization of the results of Lieb and Robinson to systems with a
power-law interaction $1/d^\alpha$ (with $d$ being the separation)
in $D$ spatial dimensions~\cite{HastingsK2006} gave a bound of $t
\sim \log d$ for the casual region of a local perturbation when
$\alpha>D$, which suggests that entanglement spreading may even
happen exponentially fast in long-ranged interacting systems. This
bound was then significantly improved in
Ref.~\onlinecite{Foss-FeigGCG2015} which applies for $\alpha>2D$ and
gives the bound for entanglement spreading as $t \sim d^\zeta$ with
$\zeta \leq 1$ and approaching $1$ as $\alpha \rightarrow \infty$
for a local perturbation. The study of quenches in different
one-dimensional models where interactions decay as a
power-law~\cite{HaukeT2013, RegemortelSW2016, BuyskikhFSED2016,
DuttaD2017, CevolaniDCTS2017} $1/d^\alpha$ shows that when $\alpha >
2$, a {\it sharp} light cone is still present in the dynamics just
like for short-range models. The light cone is
significantly broadened in the regime $1<\alpha <2$ which has been
dubbed as the {\it quasi long-range interaction regime} in Refs.\
\onlinecite{HaukeT2013, RegemortelSW2016, BuyskikhFSED2016}. For
$\alpha <1$, in contrast, the light cone effect is completely absent
with correlations between distant points building up
instantaneously. In this work, we instead focus on the entanglement
propagation for periodically driven long-ranged systems with local
quantities being observed stroboscopically (i.e. after
$n=0,1,2,\cdots$ where $n$ denotes the number of full drive cycles).
When the driving frequency $\omega$ is large, the time-evolution at
stroboscopic times can be equivalently described by a global quantum
quench where the post-quench Hamiltonian equals the time-averaged
Hamiltonian over one cycle of the periodic drive. It is then
interesting to ask whether new features that are not present for
global quenches, can emerge for the spreading of entanglement at
finite $\omega$.

Another quantity that characterizes the entanglement of a subsystem
with its environment is its entanglement entropy $S$ \cite{enrev1}.
It is defined through the reduced density matrix $\rho_r$ of the
subsystem obtained after integrating out the environment via the
following relation:
\begin{eqnarray}
S=-\mathrm{Tr} \left(\rho_r \mathrm{ln} \rho_r \right).
\label{Svn}
\end{eqnarray}
How does the entanglement entropy $S$ of the subsystem converges to
the final entanglement entropy in the steady state as a function of
time? This convergence also characterizes the approach of those
local properties that can be defined using the lattice sites
contained in the subsystem to their final steady state values since
these are fully determined by $\rho_r$. It was recently found that
the behavior of this quantity as a function of $n$ (the stroboscopic
time) shows a {\it dynamical phase transition}~\cite{SenNS2016} for
a class of integrable models in one and two dimensions, that include
the one-dimensional $S=1/2$ transverse field Ising
model~\cite{Sachdevbook} and the two-dimensional $S=1/2$ Kitaev
model~\cite{Kitaev2006}. It was shown that when a
parameter in the Hamiltonian of these models is driven periodically
in time, the local properties of the system converge to the final
steady state in two entirely different manners (which can be
identified with the two dynamical phases) depending on the driving
frequency $\omega$. However, the systems studied in
Ref.~\onlinecite{SenNS2016} have interactions whose range do not
extend beyond nearest neighbors.

In this work, we address various yet unanswered
questions regarding entanglement generation and its spreading in
periodically driven systems where the degrees of freedom are coupled
by variable range pair potentials that decay as a power law of the
form $1/d^\alpha$ with distance $d$. For instance, how does the
presence of long-ranged terms in the Hamiltonian with the range
being controlled by $\alpha$ affect the propagation of entanglement
under periodic driving? How do such systems converge to their final
nonequilibrium steady state and are there distinct dynamical phases
which are distinguished by the nature of the relaxation of local
quantities? Finally, does the light cone effect survive as a
function of $\alpha$ when the entanglement propagation is considered
stroboscopically, and do qualitatively new features emerge as a
function of $\omega$? We take a tractable model of a generalized
Kitaev chain which consists of free fermions on a one-dimensional
lattice with p-wave pairing terms that decay as $1/d^\alpha$ and
drive it periodically in time starting from an initial unentangled
pure state to address these issues.

The rest of the paper is organized in the following manner. In
Sec.~\ref{prelims}, we define the generalized Kitaev chain where the
pairing terms in the Hamiltonian are chosen to have a spatial power
law decay characterized by an exponent $\alpha$. We introduce a
pseudospin representation which allows us to express the
time-dependent Hamiltonian of the system in terms of Pauli matrices.
Using this representation, we obtain the corresponding Schrodinger
equation and solve it numerically for a specific square-pulse
periodic drive protocol characterized by a time period $T=2
\pi/\omega$ where $\omega$ is the drive frequency. In
Sec.~\ref{dynamicaltransitions}, we discuss the convergence of the
local properties of the system to their final steady state values as
a function of the number of drive cycles $n$ which plays the role of
time for stroboscopic measurement of system properties at times
$t=nT$. We identify a critical value of $\alpha=\alpha_c$, where
$\alpha_c$ depends only on the time-averaged Hamiltonian, above
which the system exhibits two dynamical phases separated by at least
one dynamical phase transition as a function of $\omega$; these
phases are distinguished by the manner in which all local
correlation functions (and hence the density matrix of a subsystem
of the system) converge to their steady state value for $n \gg 1$.
In particular, for $\omega > \omega_c^{(1)}$ (which denotes the
{{\it largest}} frequency at which the last dynamical phase
transition occurs as the frequency is varied in $[0,\infty)$), all
correlation functions shows a $n^{-3/2}$ decay to their steady state
value; this behavior changes to $n^{-1/2}$ decay as $\omega$ is
reduced through $\omega_c^{(1)}$. Such dynamical phases are
generically independent of the periodic drive protocol and show a
re-entrant behavior as a function of frequency. Below $\alpha_c$,
the high frequency dynamical phase is entirely absent and the
relaxation follows $n^{-1/2}$ behavior for any $\omega$ (apart from
some fine-tuned regions). Thus, there is a dynamical phase
transition even in the global quench limit as a function of $\alpha$
where the late-time relaxation of local properties to the steady
state changes from $t^{-3/2}$ to $t^{-1/2}$ below $\alpha_c$. We
also discuss the protocol and Hamiltonian parameter dependence of
$\alpha_c$. In Sec.~\ref{MI}, we focus on the spreading of
entanglement in the periodically driven long-ranged Kitaev chain as
a function of space and time. We show that many features of the
entanglement spreading can be understood from the behavior of the
first and second derivatives of the Floquet Hamiltonian in momentum
space. Importantly, if the decay exponent of the pairing terms is
above $\alpha_c$, we show that entanglement spreading is similar to
that of a sudden global quantum quench as long as the driving
frequency is higher than $\omega_c^{(1)}$. In contrast,
qualitatively new features emerge below $\omega_c^{(1)}$ due to {\it
additional zeroes} in the derivatives of the Floquet Hamiltonian in
momentum space. These include the appearance of multiple light
cone-like structures in the entanglement spreading in space-time
even at large $\alpha$ (i.e., effectively short-ranged models),
something which is absent for unitary dynamics after a global
quench. For $\alpha \le 2$, we also show that the entanglement
spreading is instantaneous at any drive frequency due to the
behavior of the Floquet group velocity leading to absence of light
cone like structure, which is qualitatively different from the
$\alpha>2$ case where a light cone effect exists at any drive
frequency. Finally, we discuss our main results and conclude in
Sec.~\ref{conclude}.

\section{Preliminaries}
\label{prelims}
We focus on an exactly solvable fermionic model, the generalized
Kitaev chain, with variable range p-wave pairing terms that decay as
$1/d^{\alpha}$ with the distance $d=|i-j|$ between
two lattice sites with coordinates $i$ and $j$. The Hamiltonian of
the model is as follows:
\begin{eqnarray}
H &=&-t_h\sum_{j=1}^{L}(c_{j}^{\dagger}c_{j+1}+\mathrm{H.c.})+g(t)\sum_{j}^{L}(n_{j}-1/2)\nonumber \\
&+&\frac{\Delta}{2}\sum_{j=1}^{L}\sum_{l=1}^{L-1}\left(\frac{c_{j}c_{j+l}+\mathrm{H.c.}}{d_{l}^{\alpha}}\right),
\label{Hamiltonian}
\end{eqnarray}
where  $c_{j} (c_{j}^{\dagger})$ denotes the (spinless) fermionic
annihilation (creation) operator at site $j$ and
$n_{j}=c_{j}^{\dagger}c_{j}$ is the corresponding fermion number
operator.  $t_h$ represents the fermionic hopping strength, $\Delta$
denotes the pairing between fermions, and $g(t)$ represents the
time-dependent chemical potential which is varied in a periodic
manner in time. Henceforth, we set $t_h=\Delta=1/2$. We focus on the
case of even $L$ (where $L$ denotes the number of sites in the
lattice) with antiperiodic boundary conditions for the fermions. We
accordingly define $d_{l} =l$ if $l\leq L/2$ and
$d_{l} = (L-l)$ otherwise.

When the pairing terms are restricted to be non-zero only for
nearest neighbors on the lattice, this model can be mapped via the
Jordan-Wigner transformation~\cite{Sachdevbook, LiebSM1961} to the
$S=1/2$ transverse field Ising model. The model possess two critical
points ($g=\pm 1$) in this limit and furthermore, the phase diagram
is symmetric under $g \rightarrow -g$. The correlation functions
decay exponentially in space except at the critical points. For
finite $\alpha$, the correlation functions decay exponential at
short distances but algebraically at long range for $\alpha > 1$ and
purely algebraically when $\alpha < 1$. We refer the readers to
Ref.~\onlinecite{VodolaLEGP2014} for the equilibrium phase diagram
and phase transitions of Eq.~\ref{Hamiltonian} for finite $\alpha$.

In order to diagonalize the Hamiltonian (Eq.~\ref{Hamiltonian}), we go to
the momentum space using the following transformation:
\begin{eqnarray}
c_{k}=\frac{e^{i\pi/4}}{\sqrt{L}}\sum_{x}e^{-ikx}c_{x}.
\label{Ftransform}
\end{eqnarray}
where the momenta $k$ equal $2\pi m/L$ where
$m=-(L-1)/2,....-1/2,1/2,.....(L-1)/2 $.
Writing the Hamiltonian in
terms of $c_{k}$, $c_{k}^{\dagger}$, we get
\begin{eqnarray}
H&=&\sum_{k} \Big [(g(t)-\cos(k))c_{k}^{\dagger}c_{k} \nonumber \\
&+&\Delta_{k,\alpha}(c_{-k}c_{k}+\mathrm{H.c.})-\frac{g(t)}{2} \Big]
\label {Hk1}
\end{eqnarray}
where
\begin{eqnarray}
\Delta_{k,\alpha}=\left(1/2\right)\sum_{l=1}^{L-1} \left(\sin(kl)/d_{l}^{\alpha} \right).
\label{defDeltak}
\end{eqnarray}
When $L \rightarrow \infty$, this can be written as
$\Delta_{k,\alpha}=\Im\left(\mathrm{Li}_\alpha(e^{ik})\right)$,
where $\mathrm{Li}_s(z)$ is the polylogarithm
function of order $s$ and argument $z$ and $\Im$
denotes the imaginary part of a complex number.

We note that $H_k$ connects the vacuum of the fermions $|0\rangle$
with $|k,-k\rangle$=$c_{k}^{\dagger}c_{-k}^{\dagger}|0\rangle$ and
$|k\rangle=c_{k}^{\dagger}|0\rangle$ with
$|-k\rangle=c_{-k}^{\dagger}|0\rangle$. In this work, the initial
pure state is taken to be the vacuum of the $c$ fermions. It is then
enough to consider the states $|0 \rangle, |k,-k \rangle$ at each
$k>0$ for the subsequent unitary dynamics. Furthermore, we introduce
a pseudospin representation $\vec{\sigma}_k$ where $|\uparrow
\rangle_k = |k,-k \rangle =c_k^\dagger c_{-k}^\dagger|0 \rangle$ and
$|\downarrow \rangle_k = |0\rangle$~\cite{KolodrubetzCH2012}.
Writing $H_k$ (Eq.~\ref{Hk1}) in this basis, we get
\begin{eqnarray}
H_k = (g(t)-\cos(k))\sigma_k^z + (\Delta_{k,\alpha})\sigma_k^x
\label{pseudospinH}
\end{eqnarray}

For driving protocols that preserve translational symmetry, each $k$
mode evolves independently as
\begin{eqnarray}
i\frac{d}{dt}|\psi_k \rangle=H_k(t)|\psi_k \rangle
\label{evol_k}
\end{eqnarray}
where
\begin{eqnarray}
|\psi_k (t)\rangle &=& u_k(t)|\uparrow \rangle_k + v_k(t)|\downarrow
\rangle_k , \nonumber\\
|\psi(t) \rangle &=& \otimes_{k>0}|\psi_k(t) \rangle.
 \label{product}
\end{eqnarray}

Thus, specifying $u_k,v_k$ for $k>0$ specifies the complete wavefunction
of the system through Eq.~\ref{product}. The initial
state can be easily expressed in the pseudospin basis as
$|\psi(0) \rangle = \otimes_{k>0} |\downarrow \rangle_k$.

For numerical convenience, we take the time-dependence of $g(t)$ as a square
pulse that varies periodically in time with a period that equals $T$, i.e.,
\begin{eqnarray}
g(t)&=& g_i, \quad (n-1)T \le t\le (n-1/2)T \nonumber\\
&=& g_f, \quad (n-1/2)T \le t\le n T,
\label {squarepulse}
\end{eqnarray}

Since we are interested in the stroboscopic behavior of the local quantities,
it is enough to know the unitary time evolution operator $U_k(T)$
at each $k$ for a single period $T$.
The unitary evolution after a time $t=nT$ where $n=0,1,2,\cdots$ can be
calculated as
\begin{eqnarray}
|\psi_k (nT)\rangle = [U_k(T)]^n|\psi_k(0) \rangle.
\label{strobevol}
\end{eqnarray}

We note here that most of our results are independent of the specific
form of the periodic drive protocol and the above protocol has been taken
to make the analysis tractable.
\section{Convergence to the steady state and dynamical phase transition}
\label{dynamicaltransitions}

In this section, we discuss the convergence of the local properties
of the generalized Kitaev chain (Eq.~\ref{Hamiltonian}) when $g(t)$
is driven periodically in time. For this, we will use the formalism
developed in our earlier work~\cite{SenNS2016} in the context of
short-ranged integrable models with no interactions beyond nearest
neighbors and show how it generalizes to the present case where the
pairing terms decay as a power law in space.

Since we are dealing with a quadratic fermionic Hamiltonian in
Eq.~\ref{Hamiltonian}, it is enough to consider the behavior of the
two-point correlators $C_{ij}(n)=\langle c_i^\dagger c_j \rangle_n$
and $F_{ij}(n)=\langle c_i^\dagger c_j^\dagger \rangle_n$
stroboscopically (i.e., at $t=nT$) to study the convergence to a
possible final steady state as $n \rightarrow \infty$. Other
higher-point correlators can then be constructed from $C_{ij}(n)$
and $F_{ij}(n)$ by using Wick's theorem. It is useful to look at
this problem for a general periodic drive protocol that preserves
the lattice translational symmetry first.

Eq.~\ref{pseudospinH} describes the motion of the pseudospin
$\vec{\sigma}_k$ at momentum $k$ in a time-varying ``magnetic
field'' $(\Delta_{k,\alpha},0,g(t)-\cos(k))$. The time evolution
operator for one time period for $\vec{\sigma}_k$ can thus be
parametrized as $U_k(T) =\exp[-iH_{kF}T]$ where the hermitian
operator $H_{kF}$ is the Floquet Hamiltonian of the system at
momentum $k$, which can be written in general as
\begin{eqnarray}
H_{kF} = \vec{\sigma}_k \cdot \vec{\epsilon}_{k}=|\vec{\epsilon}_{k}|\vec{\sigma}_k\cdot \hat{n}_k
\label{HkF}
\end{eqnarray}
where $\vec{\epsilon}_{k} = (\epsilon_{k1},
\epsilon_{k2}, \epsilon_{k3})$, and
$\hat{n}_{ki}=\epsilon_{ki}/|\vec{\epsilon}_{k}|$. Then, we can
express $U_k(T)$ as
\begin{eqnarray}
U_k(T)=\exp[-i(\vec{\sigma}_k \cdot \hat{n}_k) \phi_k]
\label{Uform}
\end{eqnarray}
where $\phi_k=T|\vec{\epsilon}_{k}|$ and we restrict $\phi_k \in
[0,\pi]$ and each component of $\vec{\epsilon}_{k} \in
[-\pi/T,\pi/T]$ without loss of generality ({\it
i.e.}, we use the reduced zone scheme).

We now study the behavior of $C_{ij}$ and $F_{ij}$
stroboscopically when the initial pure state is taken to be the
vacuum of the fermions, i.e., $u_k(0)=0$ and $v_k(0) =1$ for all $k$.
Using the form of $|\psi(t) \rangle$ in Eq.~\ref{product}, we get
\begin{eqnarray}
C_{ij}(n) &=& \langle c_i^\dagger c_j \rangle_n =\frac{2}{L} \sum_{k>0}|u_k(nT)|^2 \cos(k(i-j)) \label{cfdef} \\
F_{ij}(n) &=& \langle c_i^\dagger c_j^\dagger \rangle_n =\frac{2}{L} \sum_{k>0}u_k^*(nT)v_k(nT) \sin(k(i-j)) \nonumber
\end{eqnarray}
Using Eq.~\ref{strobevol} and Eq.~\ref{Uform}, and taking the $L
\rightarrow \infty$ limit in Eq.~\ref{cfdef}, we get the following
expressions for $\delta C_{ij}(n)= C_{ij}(n)-C_{ij}(\infty)$ and
$\delta F_{ij}(n) = F_{ij}(n)-F_{ij}(\infty)$, where
$C_{ij}(\infty)$ and $F_{ij}(\infty)$ are the steady state values of
the correlators: \cite{SenNS2016}
\begin{eqnarray}
\delta C_{ij}(n) &=& \frac{1}{\pi} \int_0^\pi dk \mathcal{F}_1(k)\cos(2n\phi_k) \nonumber \\
\delta F_{ij}(n) &=& \frac{1}{\pi} \int_0^\pi dk [\mathcal{F}_2(k)
\cos(2n\phi_k) + \mathcal{F}_3\sin(2n \phi_k)]
\nonumber\\
C_{ij}(\infty) &=& \frac{1}{\pi} \int_0^\pi dk \cos(k(i-j))
\left(\frac{1}{2}(1-\hat{n}_{k3}^2) \right)  \label{cfijexp} \\
F_{ij}(\infty) &=& \frac{1}{\pi} \int_0^\pi dk \sin(k(i-j))
\left(-\frac{1}{2}\hat{n}_{k3}(\hat{n}_{k1}+i\hat{n}_{k2}) \right)
\nonumber
\end{eqnarray}
where
\begin{eqnarray}
\mathcal{F}_1(k) &=& -\frac{1}{2}\cos(k(i-j))(1-\hat{n}_{k3}^2), \mathcal{F}_2(k)=-i\hat{n}_{k3}\mathcal{F}_3(k) \nonumber \\
\mathcal{F}_3(k) &=& \frac{i}{2}\sin(k(i-j))
(\hat{n}_{k1}+i\hat{n}_{k2}). \label{F123}
\end{eqnarray}
We note that while converting the summation over $k$ (in
Eq.~\ref{cfdef}) to an integral (in Eq.~\ref{cfijexp}), we have
implicitly assumed that $|i-j| \ll L$. The steady state is strictly
reached only for such local operators (where $|i-j| \ll L$) and
$C_{ij}(n) (F_{ij}(n))$ continues to display undamped oscillations
even when $n \rightarrow \infty$ if $|i-j| \sim \mathcal{O}(L)$.

From Eq.~\ref{cfijexp}, it is clear that $\delta C_{ij}(n)$ and
$\delta F_{ij}(n)$ must vanish when $n \rightarrow \infty$ by the
Riemann-Lebesgue lemma. Moreover, the dominant contribution to this
relaxation behavior to the steady state is controlled by the
stationary points defined by $d|\vec{\epsilon}_{k}|/dk=0$ at late
times. The contribution of such a stationary point at $k=k_0$ to
$\delta C_{ij}(n)$ and $\delta F_{ij}(n)$ can be estimated using the
stationary phase approximation:\cite{SenNS2016}
\begin{eqnarray}
\int \mathcal{F}_i(k)\exp(in\phi_k)dk \approx \exp(in\phi_{k_0})(n|\phi^{''}(k_0)|)^{-1/2} \nonumber \\
\times \exp \left(\frac{\pi i \mu}{4} \right) \left(\mathcal{F}_i(k_0)+i\frac{\mathcal{F}_i^{''}(k_0)}{2\phi^{''}(k_0)} \frac{1}{n} +\mathcal{O} \left(\frac{1}{n^2} \right)\right)
\label{stationarypt}
\end{eqnarray}
where $\mu$ is the sign of $\phi^{''}(k_0)$ and $\mathcal{F}_i(k)$
is assumed to be a smooth function in the neighborhood around
$k=k_0$.

Importantly, possible stationary points at the Brillouin zone (BZ)
edges, $k=0$ and $k=\pi$, behave differently to those where $k \in
(0,\pi)$ (i.e., excluding $k=0$ and $k=\pi$).~\cite{SenNS2016} To
see this, we first note that $\Delta_{k,\alpha}=-\Delta_{-k,\alpha}$
for the long-ranged Kitaev chain independent of the
value of $\alpha$. Thus, $\Delta_{k,\alpha}=0$ at the edges of the
BZ. From this, it follows that $U_k(T)$ is diagonal in the
$|\uparrow \rangle_k,|\downarrow \rangle_k$ basis at $k=0,\pi$ and
hence $\hat{n}_{k1}=\hat{n}_{k2}=0$ and $\hat{n}_{k3}=\pm 1$ for any
periodic drive protocol. From Eq.~\ref{F123}, it then follows that
$\mathcal{F}_{1,2,3}(k_0)=0$ for $k_0=0,\pi$; in contrast, for $k_0
\ne 0, \pi$, they are in general non-zero. Using this result, from
Eq.\ \ref{stationarypt}, it is easy to see that the stationary
points at the edges of the BZ thus lead to $\mathcal{O}(n^{-3/2})$
decay of the correlation functions (Eq.\
 \ref{cfijexp}); in contrast, for $k_0 \ne 0,\pi$, the decay is
$\mathcal{O}(n^{-1/2})$. Since $\mathcal{F}_{1,2,3}(k)=0$ both at
$k=0$ and $k=\pi$, in the absence of any stationary points in $k \in
[0,\pi]$, $\delta C_{ij}(n)$ ($\delta F_{ij}(n)$) would have decayed
as $\mathcal{O}(n^{-2})$ (and not as $\mathcal{O}(n^{-1})$ which
requires $\mathcal{F}_{1,2,3}(k) \neq 0$ at least at one of the BZ
edges) which is sub-leading compared to both $n^{-1/2}$ and
$n^{-3/2}$ decays as $n \gg 1$.

To study the relaxation of the entanglement entropy of a subsystem
of $l$ sites (which are assumed to be adjacent for concreteness) to
its final steady state value $S_\infty(l)$, we note that for a
Hamiltonian of the form Eq.~\ref{Hamiltonian}, both the reduced
density matrix $\rho_r$ of the subsystem and its entanglement
entropy $S_n(l)$ with the environment may be calculated from the
knowledge of $C_{ij}(n)$ and $F_{ij}(n)$ where $i,j$ denote the
sites that belong to the subsystem.~\cite{VidalLRK2003, Peschel2003}
Two $l \times l$ matrices can be constructed from $C_{ij}(n)$ and
$F_{ij}(n)$, which we denote by $\mathbf{C}$ and $\mathbf{F}$
respectively. From these, we construct the following $2l \times 2l$
matrix:
\begin{eqnarray}
{\mathcal C}_n(l) &=& \left( \begin{array}{cc} \mathbf{I-C} &
\mathbf{F} \\ \mathbf{F}^{\ast} & \mathbf{C } \end{array} \right).
\label{smatrix}
\end{eqnarray}
$S_n(l)$ can then be obtained from the $2l$ eigenvalues (denoted by $p_i$)
of the matrix $\mathcal{C}_n(l)$: $S_n(l)= -\mathrm{Tr}[\rho_r \ln \rho_r] =
-\sum_{i=1}^{2l}p_i \ln(p_i)$. Furthermore, $\rho_r$ can be obtained by
knowing the eigenvectors of $\mathcal{C}_n(l)$ as well.

To characterize the approach of $\rho_r$ to the final reduced
density matrix of the steady state, we define the following distance
measure~\cite{FagottiE2013} $\mathcal{D}_n(l)$:
\begin{eqnarray}
\mathcal{D}_n(l) = \mathrm{Tr}[(\mathcal{C}_{\infty}(l)-\mathcal{C}_n(l))^\dagger (\mathcal{C}_{\infty}(l)-\mathcal{C}_n(l))]^{1/2}/(2l).
\label{distancemeasure}
\end{eqnarray}
This distance measure has the property that $0 \leq \mathcal{D}_n(l) \leq
1$ and vanishes only when $\mathcal{C}_n(l)=\mathcal{C}_\infty(l)$,
which also implies that $\rho_r$ itself has converged to the final
steady state reduced density matrix for the subsystem. From the
discussion on stationary points above, we thus see that if such
stationary points are solely present on the edges of the BZ, then
all the elements of $\mathcal{C}_n(l)$ and hence $\mathcal{D}_n(l)$
converge to the final steady state as $(\omega/n)^{3/2}$, while if
there are any stationary points for $k_0 \in (0,\pi)$, then the
relaxation instead shows a $(\omega/n)^{1/2}$ behavior. Thus, the
long-time relaxation properties are again controlled by whether the
number of stationary points of $|\vec{\epsilon}_{k}|$ (defined by
$d|\vec{\epsilon}_{k}|/dk=0$) inside the BZ ($0 < k < \pi$), which
we denote by $N_s$ henceforth, equals zero or not, just like in the
case of short-ranged integrable models considered in
Ref.~\onlinecite{SenNS2016}.

\begin{figure}[H]
\begin{center}
       \subfigure[]{%
            \label{fig:first}
            \includegraphics[width=0.5\textwidth]{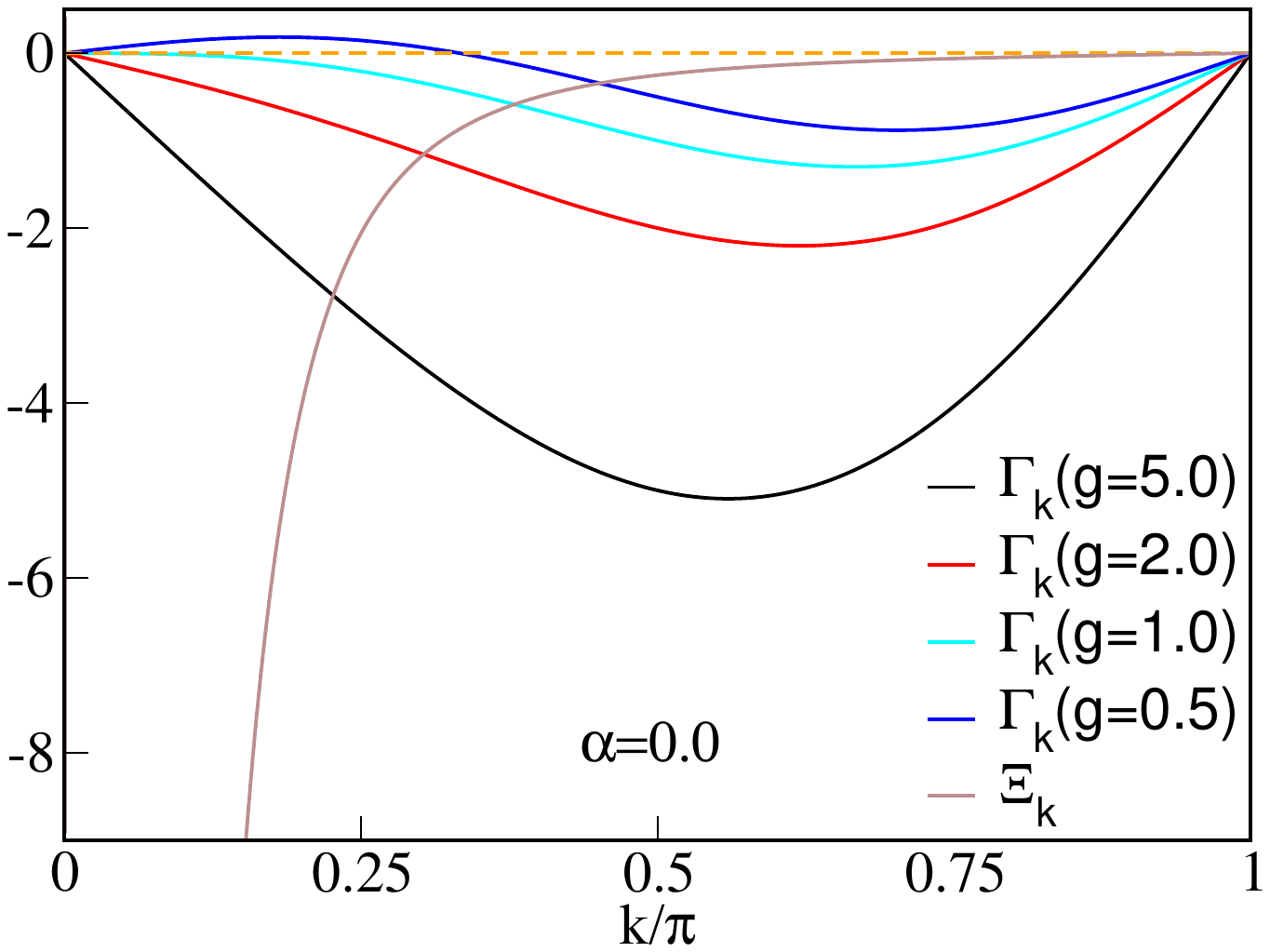}
        } \\
        \subfigure[]{%
           \label{fig:second}
           \includegraphics[width=0.5\textwidth]{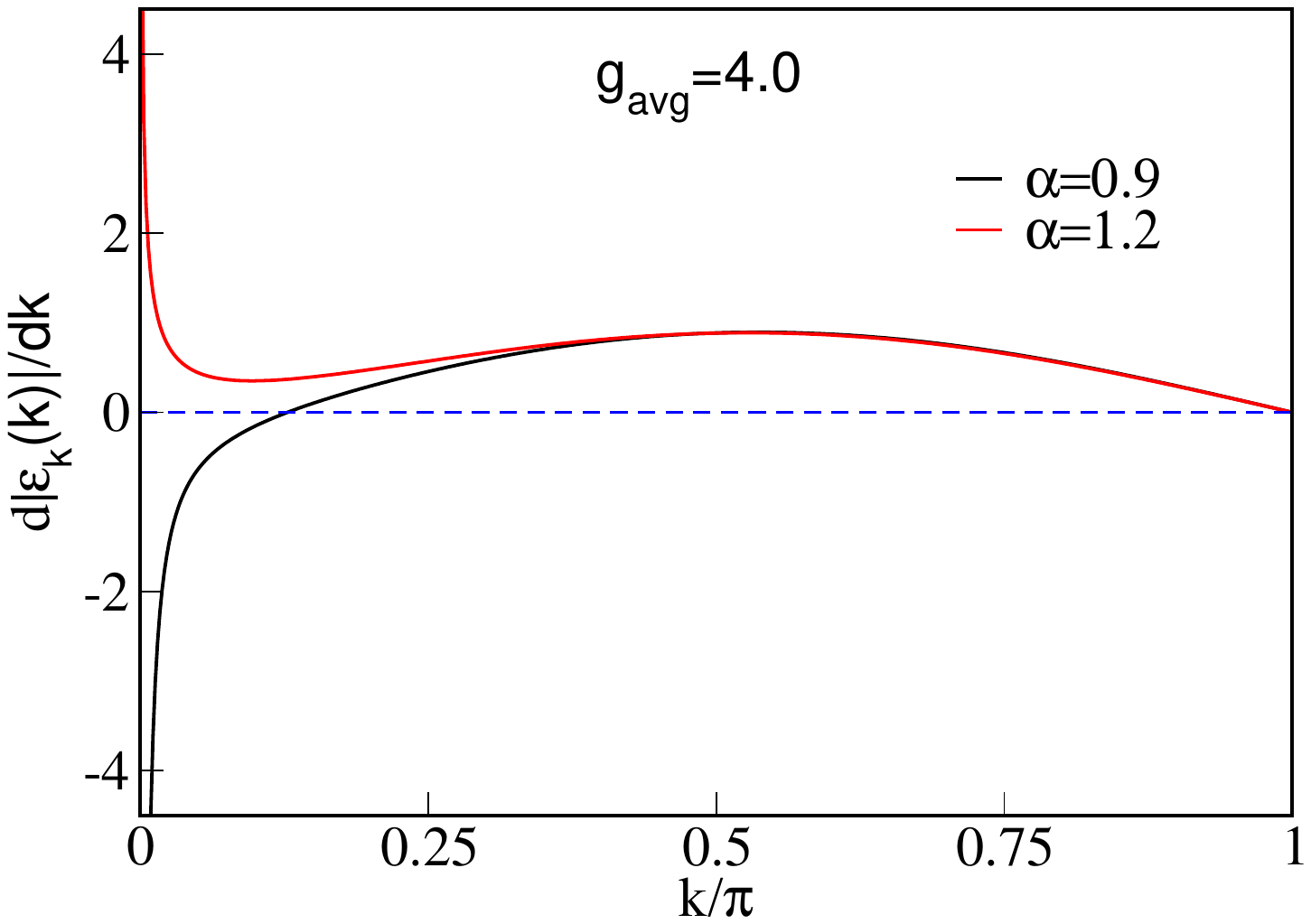}
        }  
    \end{center}
    \caption{%
 (a) The behaviour of the functions $\Gamma_k(g_{avg})$ and $\Xi_k$ as
a function of $k$. (b) The behavior of $d|\vec{\epsilon}_{k}|/dk$
shown as a function of $k$
at $g_{avg}=4$ for two particular values of $\alpha$.
For $\alpha = 1.2$ (red curve),
the stationary point is only present at $k=\pi$, while for $\alpha = 1.2$ (black
curve), an additional stationary point is present in $k \in (0,\pi)$.}%
   \label{fig1}
\end{figure}

\begin{figure}[H]
\centering
{\includegraphics[width=\hsize]{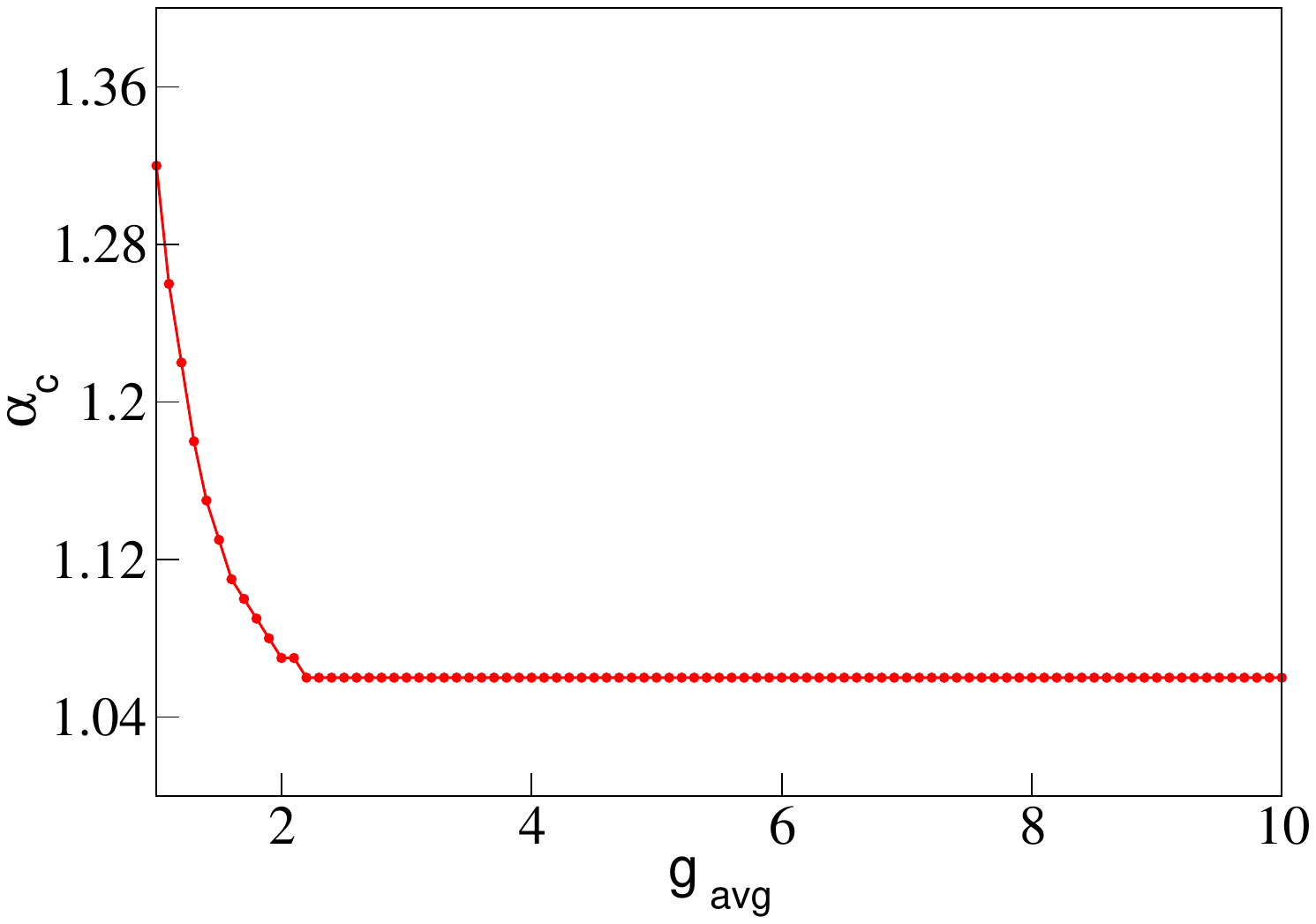}}
\caption{The behavior of $\alpha_c$ as a function of $g_{avg}$.}
\label{fig2}
\end{figure}

{\it High frequency limit:} First, let us consider the case when
$\omega \rightarrow \infty$. In this limit, $H_{kF} \sim \bar{H}_k$,
where $\bar{H}$ denotes the time-averaged Hamiltonian over one drive
cycle, by using $1/\omega$ as a perturbation parameter in the Dyson
series for $U_k(T)$. $\bar{H}_k$ can be obtained from
Eq.~\ref{pseudospinH} by replacing $g(t)$ by
$g_{avg}=(1/T)\int_0^Tg(t)dt$. It then follows that
\begin{eqnarray}
|\vec{\epsilon}_{k}|_{\omega \rightarrow \infty}=\sqrt{(g_{avg}-\cos(k))^2+\Delta_{k,\alpha}^2}.
\label{largew}
\end{eqnarray}
When $\alpha \rightarrow \infty$, we see that $\Delta_{k,\alpha}
\rightarrow \sin(k)$ from Eq.~\ref{defDeltak}, from which it is
straightforward to show that the only stationary
points of $|\vec{\epsilon}_{k}|$ are at the BZ edges ($k=0,\pi$ for
$g_{avg} \neq \pm 1$, $k=0$ for $g_{avg}=-1$, and $k=\pi$ for
$g_{avg}=+1$). Next, we consider the opposite limit where $\alpha
\rightarrow 0$. It can then be shown there always exists one more
stationary point in $0<k<\pi$ by considering the behaviour of the
following functions:
\begin{eqnarray}
\Gamma_k(g_{avg})&=&(\cos(k)-g_{avg})\sin(k) \nonumber \\
\Xi_k &=& \mathrm{Lim}_{\alpha \rightarrow 0} \Im\left(\mathrm{Li}_\alpha(e^{ik})\right) \Re\left(\mathrm{Li}_{\alpha-1}(e^{ik})\right)
\end{eqnarray}
where $\Re$ ($\Im$) denotes the real (imaginary) part of a complex
number. A stationary point in $0<k<\pi$ implies that
$\Gamma_{k_0}(g_{avg})= \Xi_{k_0}$ for some $k_0 \in (0,\pi)$. This
is always guaranteed when $\alpha \rightarrow 0$ because $\Xi_k$ is
a monotonic function whose range extends from $(-\infty,0]$ in $k
\in [0,\pi]$ and $\Gamma_k(g_{avg})=0$ at the BZ edges and is
negative for $k \rightarrow \pi^-$ independent of $g_{avg}$
(Fig.~\ref{fig1}(a)). We have numerically checked that there are
stationary points only at the edges of the BZ for all $\alpha >
\alpha_c(g_{avg})$, while below $\alpha_c(g_{avg})$, additional
stationary points arise in $k \in (0,\pi)$ (Fig.~\ref{fig1}(b)). The
behavior of $\alpha_c$ as a function of $g_{avg}$ is shown in
Fig.~\ref{fig2}. We find numerically that $\alpha_c$ is constant
as a function of $g_{avg}$ ($\alpha_c \approx 1.05$) for all
$g_{avg}$ until $g_{avg} \approx 2$ and it starts to then increase
with decreasing $g_{avg}$ thereafter.

 The determination of $\alpha_c$ is completely
independent of any specific periodic drive protocol and only depends
on $g_{avg}$ since it is fixed by the behavior at $\omega
\rightarrow \infty$. We have thus unearthed a dynamical phase
transition even in the global quench limit for such long-ranged
models where the (long time) approach of local quantities to their
steady state values change from $t^{-3/2}$ for $\alpha
> \alpha(g_{avg})$ to $t^{-1/2}$ for $\alpha < \alpha_c(g_{avg})$
where the post-quench Hamiltonian's (Eq.~\ref{Hamiltonian}) chemical
potential $g$ is fixed to be $g_{avg}$.

\begin{figure}[H]
\begin{center}
       \subfigure[]{%
            \label{fig:first}
            \includegraphics[width=0.5\textwidth]{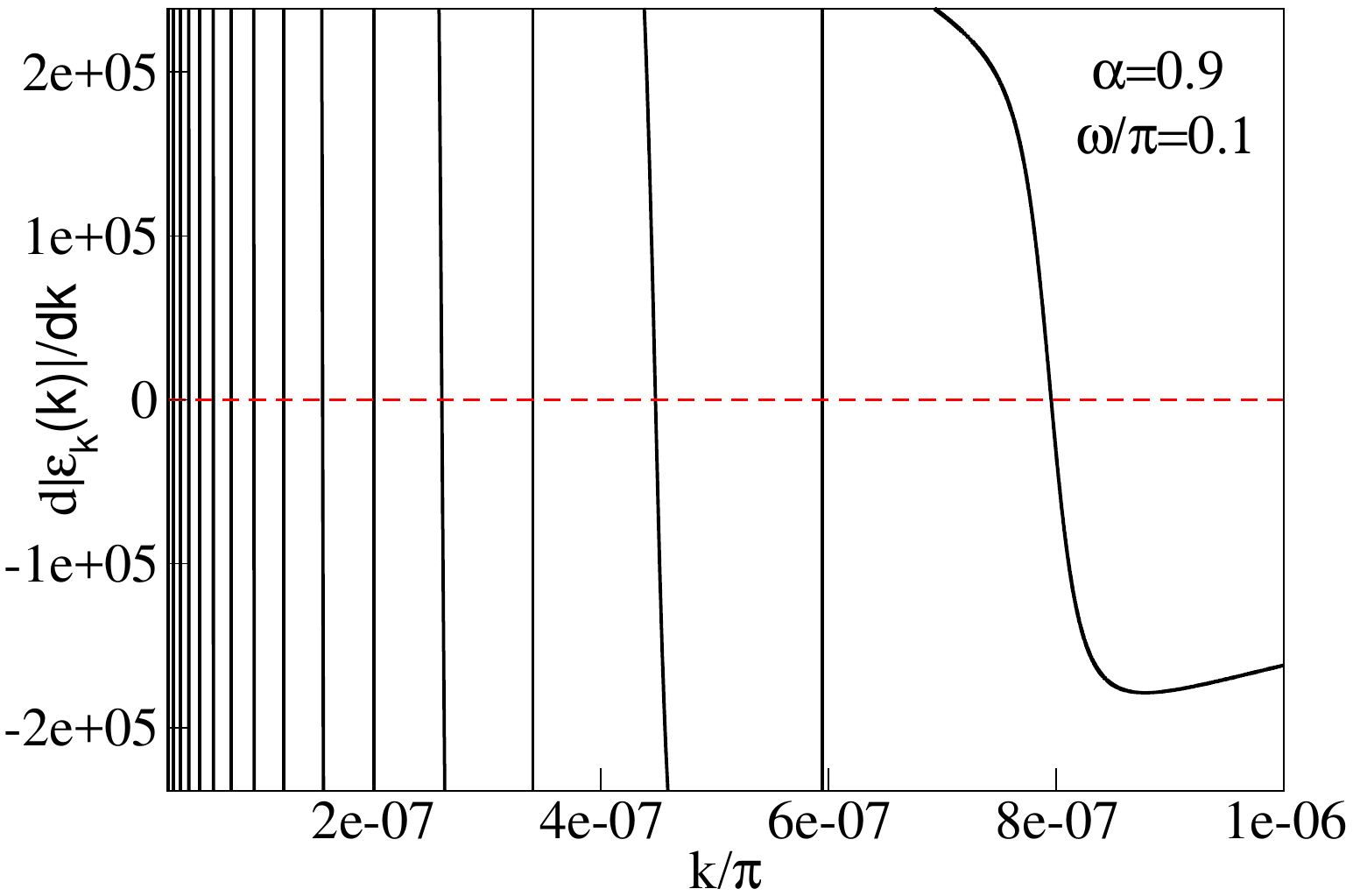}
        } \\
        \subfigure[]{%
           \label{fig:second}
           \includegraphics[width=0.5\textwidth]{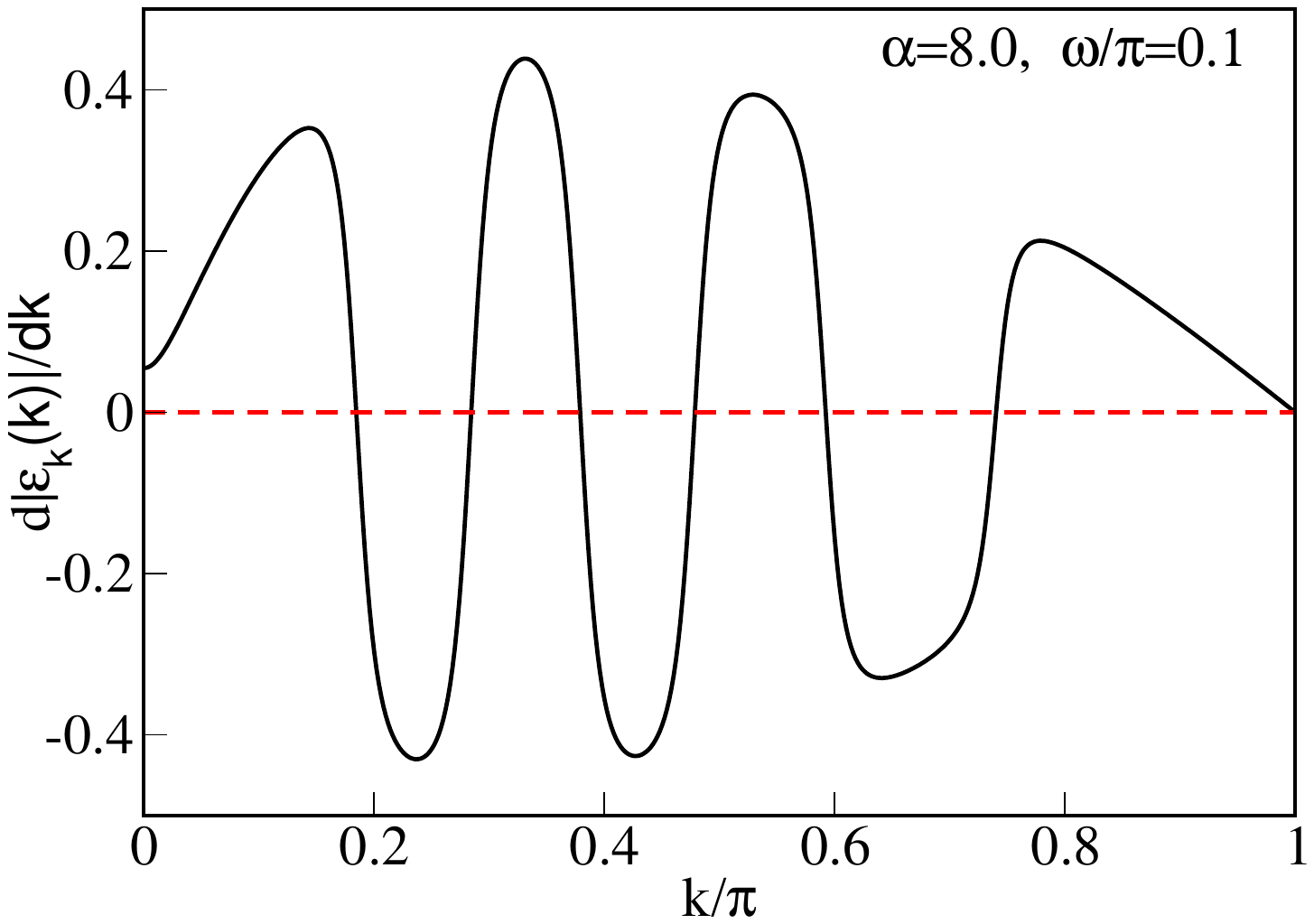}
        }  
    \end{center}

\caption{(a) The behavior of $d|\vec{\epsilon}_k|/dk$ as a function of
$k$ at a small $\omega$ ($\omega/\pi=0.1$) for $\alpha=0.9$ shows a
large number of stationary points in the vicinity of $k=0$.
(b) The behavior of $d|\vec{\epsilon}_k|/dk$ as a function of $k$ at the
same $\omega$ but at a larger $\alpha=8.0$. The drive protocol used here is
the square pulse protocol with $g_i=2$ and $g_f=0$.}
\label{fig3}
\end{figure}

\begin{figure}[H]
\centering {\includegraphics[width=\hsize]{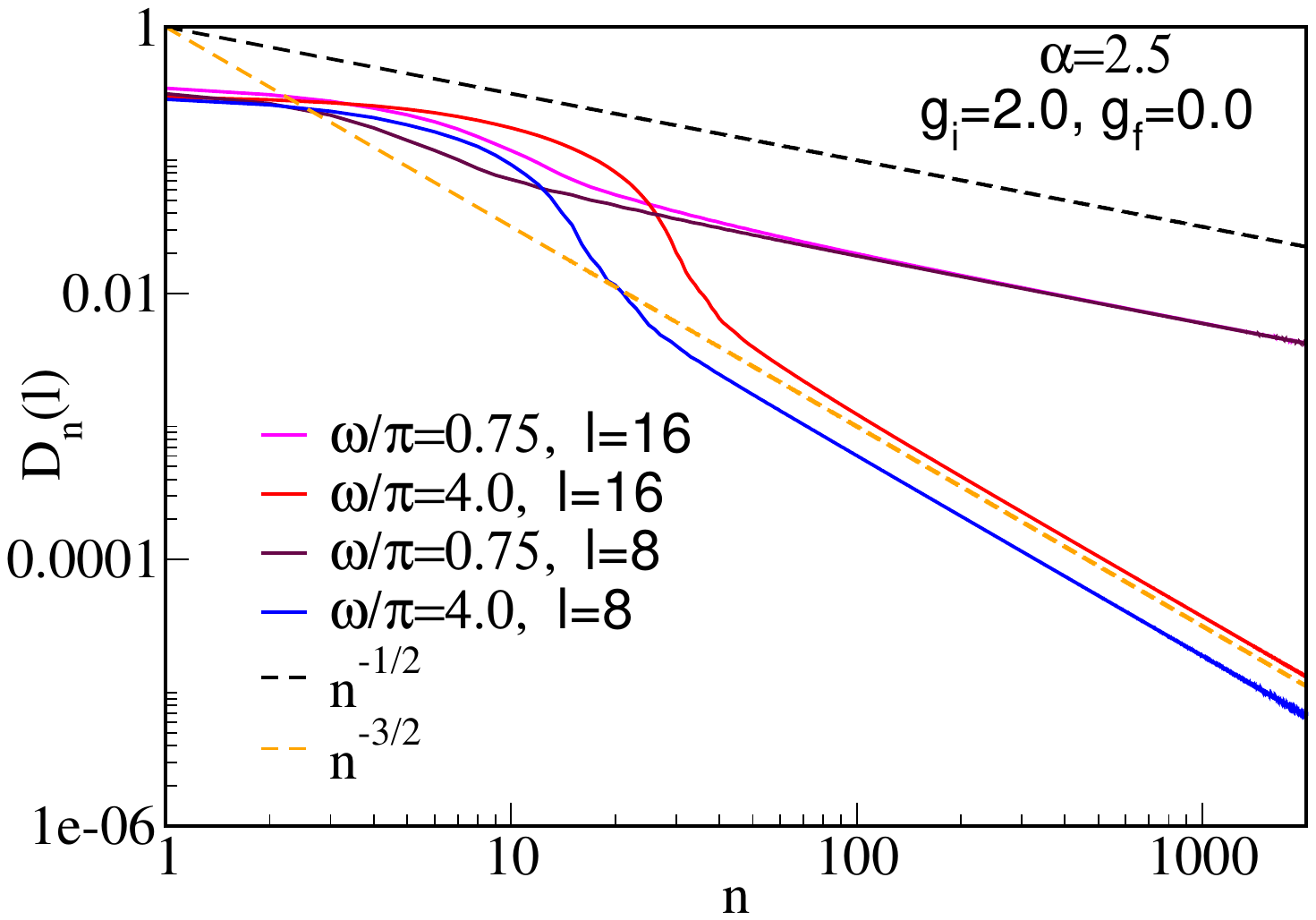}}
\caption{The behavior of $\mathcal{D}_n(l)$ as a function of $n$ for
two different driving frequencies that belong to different dynamical
phases. At $\omega/\pi=4.0 (0.75)$, $\mathcal{D}_n(l) \sim
(\omega/n)^{3/2} ((\omega/n)^{1/2})$ both for $l=8$ and $l=16$, where $l$
denotes the number of consecutive sites in the subsystem. The other
parameters are $g_i=2, g_f=0, \alpha=2.5$ and $L=2\times 10^5$.  }
\label{fig4}
\end{figure}

{\it Low frequency limit:}
Next, we discuss the behavior of $|\vec{\epsilon}_k|$ at small $\omega$.
For calculational purposes, we adopt the square pulse protocol given in
Eq.~\ref{squarepulse}. It can then be shown that \cite{SenNS2016}
\begin{eqnarray}
|\vec{\epsilon}_k|=\arccos(M_k)/T
\label{spp1}
\end{eqnarray}
where
\begin{eqnarray}
M_k = \cos(\Phi_{ki})\cos(\Phi_{kf})- \vec{N}_{ki}\cdot \vec{N}_{kf} \sin(\Phi_{ki})\sin(\Phi_{kf}). \nonumber \\
\label{spp2}
\end{eqnarray}
In the above expression, $\Phi_{ki(f)}=E_{ki(f)}T/2$ with
$E_{ki(f)}=\sqrt{(g_{i(f)}-\cos(k))^2+\Delta_{k,\alpha}^2}$ and
$\hat{N}_{ki(f)} = \left(\Delta_{k,\alpha}/E_{ki(f)},0,
(g_{i(f)}-\cos(k))/E_{ki(f)} \right)$. For large $T=2\pi/\omega$,
$M_k$ rapidly oscillates in $[-1,1]$ with the effective wavelength
being set by $1/T$ in $k$ space. Thus, when $\omega \rightarrow 0$,
the number of stationary points $N_s$ of $|\vec{\epsilon}_k|$ in
$0<k<\pi$ increases with decreasing $\omega$. In fact, we see that
$N_s \rightarrow \infty$ as $\omega \rightarrow 0$ irrespective of
the value of $\alpha$. A scaling of $N_s \sim 1/\omega$ at small
$\omega$ was previously seen in Ref.~\onlinecite{SenNS2016} for one
dimensional short-ranged integrable models. Interestingly,
decreasing the value of $\alpha$ below a certain threshold ($\alpha
\sim 1$) increases the number of stationary points greatly,
particularly in the neighborhood of $k=0$ (Fig.~\ref{fig3}(a)); an
effect which is absent for larger values of $\alpha$ as shown in
Fig.~\ref{fig3}(b). Thus the scaling of $N_s$ is actually faster
than $1/\omega$ at small $\omega$ when $\alpha$ is small.
For $\alpha>\alpha_c(g_{avg})$, we thus see that $N_s=0$ when
$\omega \rightarrow \infty$ and $N_s \sim 1/\omega$ for $\omega
\rightarrow 0$. Hence, $\mathcal{D}_l(n) \sim (\omega/n)^{3/2}$ for
fast drives and $\mathcal{D}_l(n) \sim (\omega/n)^{1/2}$ for slow
drives as long as $\alpha > \alpha_c(g_{avg})$ irrespective of the
specific details of the periodic drive protocol. As a result, there
must be at least one dynamical phase transition between these two
dynamical phases distinguished by the relaxation of
$\mathcal{D}_n(l)$ (as defined in Eq.~\ref{distancemeasure}).
Consequently {\it all} local quantities relax to their steady state
values either as $(\omega/n)^{3/2}$ or as $(\omega/n)^{1/2}$ as the
drive frequency $\omega$ is varied keeping other parameters fixed.
We illustrate this in Fig.~\ref{fig4} where $\alpha$ is taken to be
greater than $\alpha_c$. The two different drive frequencies
$\omega$ show the different scalings of $\mathcal{D}_n(l) \sim
(\omega/n)^{3/2}$ and $\mathcal{D}_n(l) \sim (\omega/n)^{1/2}$
respectively.

\begin{figure}[H]
\centering {\includegraphics[width=\hsize]{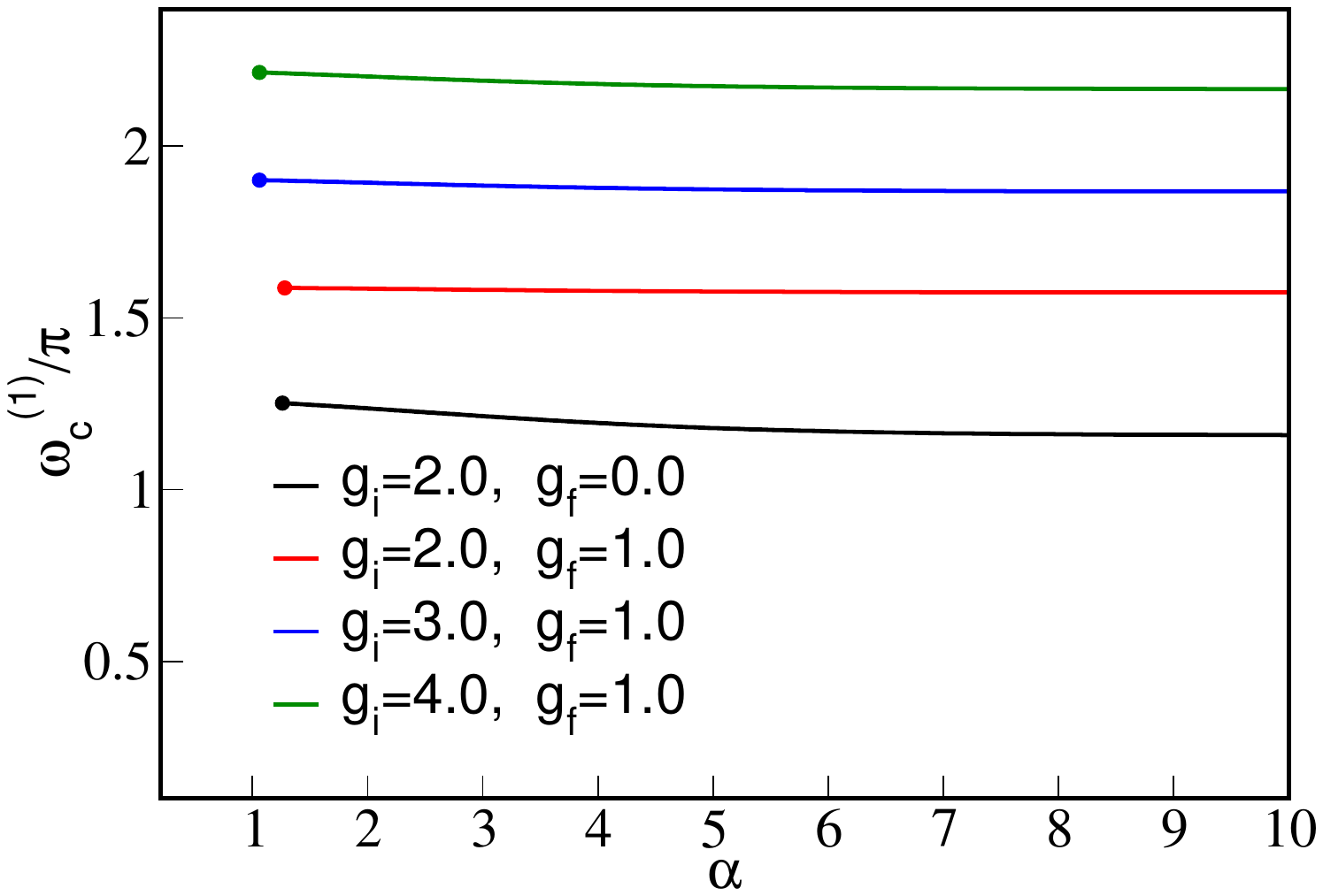}}
\caption{The variation of $\omega^{(1)}_c$ as a function of $\alpha$
for a given $g_i$ and $g_f$.}
\label{fig5}
\end{figure}

{\it Dynamical phase transitions:} Since $N_s$ is an integer, its
value {\it cannot} change smoothly from $N_s=0$ to $N_s=1$ as
$\omega$ is decreased from $1/\omega=0$ if the limit $N_s(\omega
\rightarrow \infty)$ exists, and can only turn
non-zero for the first time at a finite value of $\omega^{(1)}_c$
for any periodic drive protocol where the range of the pairing terms
is greater than $\alpha(g_{avg})$. For any $\omega \in
(\omega^{(1)}_c, \infty)$, $\mathcal{D}_l(n) \sim (\omega/n)^{3/2}$
from the previous discussion. To calculate $\omega^{(1)}_c$ for the
square pulse protocol~\cite{SenNS2016}, we note that the new zero in
$d|\vec{\epsilon}_k|/dk$ can appear only from the BZ boundaries. We
have numerically checked that irrespective of the value of $\alpha$
($>\alpha_c(g_{avg})$), the new zero emerges from $k=\pi$ for this
specific protocol. Then,
for a given $\alpha$, $\omega_c^{(1)}$ can be calculated by
expanding $d|\epsilon_{k}|/dk$ for $k=\pi-\epsilon$ and finding the
value of $\omega$ where the $\mathcal{O}(\epsilon)$ term first
changes its sign. In fact, $\omega^{(1)}_c=2\pi/T_0$ where $T_0$ is
the smallest non-zero solution of the following equation:
\begin{eqnarray}
&& 2\sin(\mathcal{G}_1T_0)[(1+g_f)^2(1+g_i)^2T_0] \nonumber\\
&& +4^{-\alpha}(2^{\alpha}-4)^2(\zeta(\alpha-1))^2 \nonumber\\
&& \times [(g_f-g_i)^2\{\cos(\mathcal{G}_2T_0)-\cos(\mathcal{G}_1T_0)\}\nonumber\\
&&-2(1+g_f)(1+g_i)\mathcal{G}_1T_0\sin(\mathcal{G}_1T_0)] =0
\label{condition}
\end{eqnarray}
where $\mathcal{G}_1=(g_i+g_f+2)/2$, $\mathcal{G}_2=(g_f-g_i)/2$ and
$\zeta(s)$ denotes the Riemann zeta function.
In Fig.~\ref{fig5}, we show how this $\omega_c^{(1)}$
varies as a function of $\alpha$ for different values of $g_i$ and $g_f$.
Interestingly, one can see that for a given set of $g_i$ and $g_f$,
$\omega^{(1)}_c$ is rather insensitive to
the variation of $\alpha$ (note that $\omega^{(1)}_c$ ceases to exist
below $\alpha(g_{avg})$ which explain the ``end-points'' in Fig.~\ref{fig5}).

\begin{figure}[H]
\begin{center}
       \subfigure[]{%
            \label{fig:first}
            \includegraphics[width=0.5\textwidth]{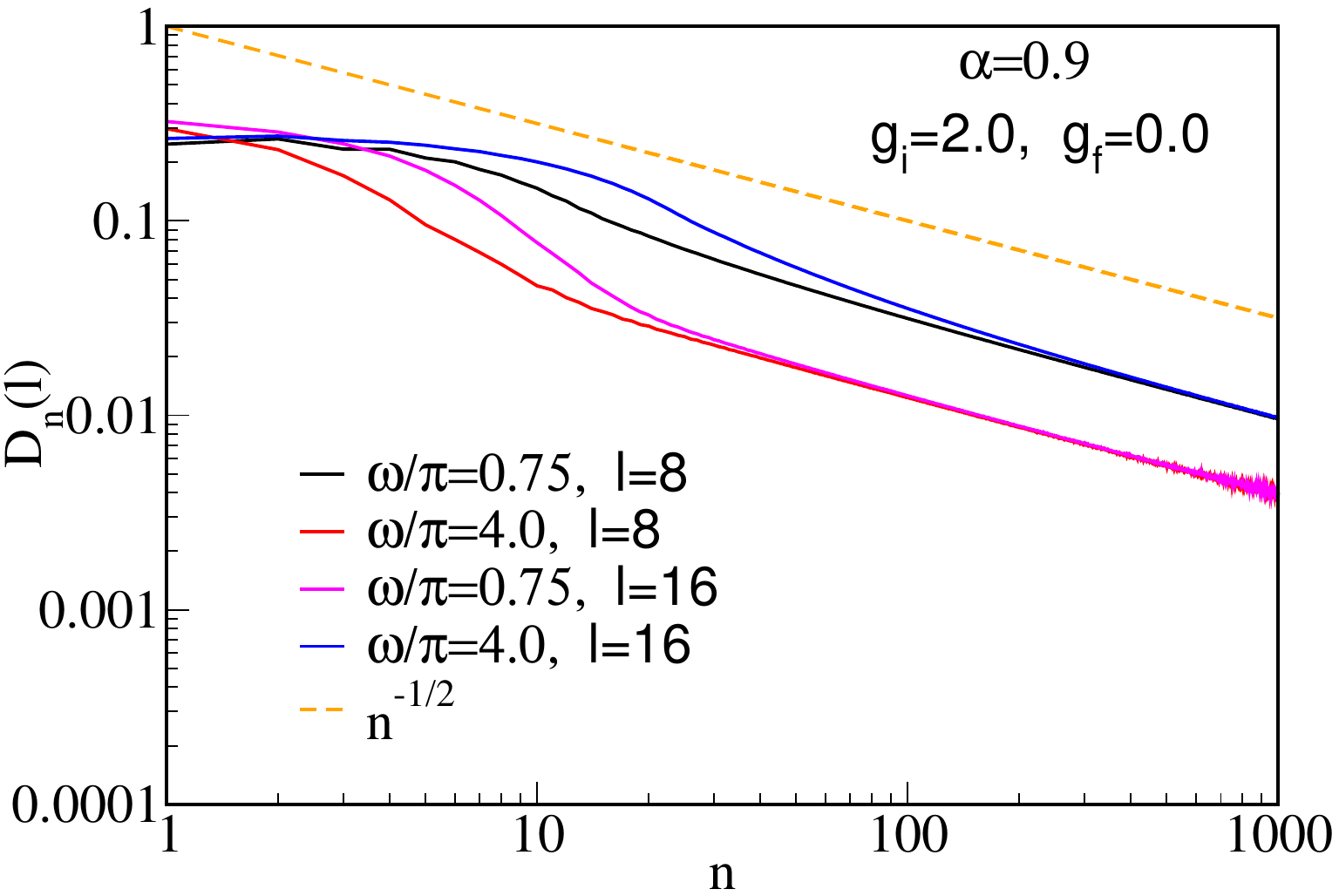}
        } \\
        \subfigure[]{%
           \label{fig:second}
           \includegraphics[width=0.5\textwidth]{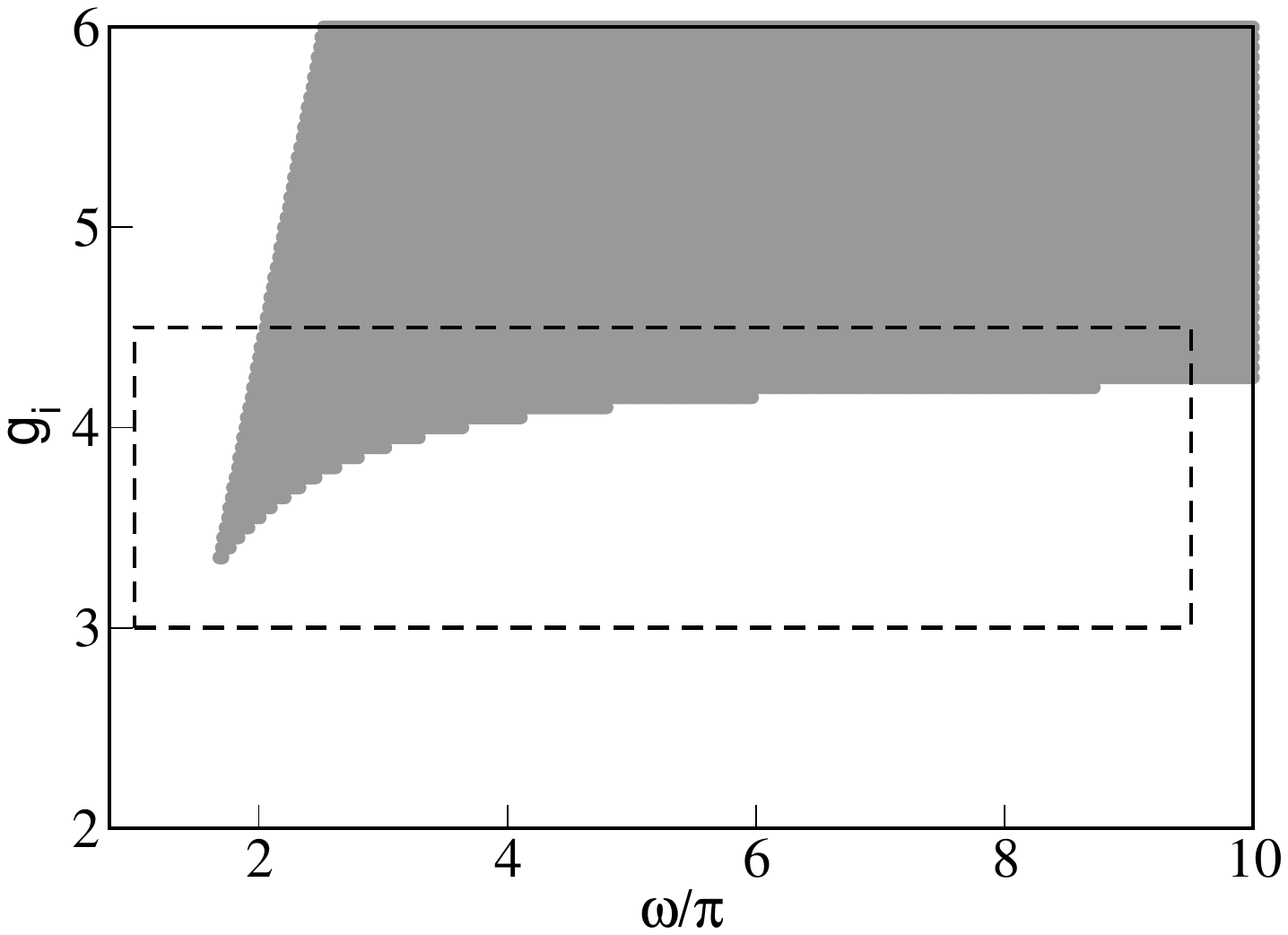}
        }   
    \end{center}
    \caption{%
 (a) The behavior of $\mathcal{D}_n(l)$ as a function of $n$ for
fast ($\omega/\pi=4.0$) and slow ($\omega/\pi=0.75$) driving frequencies
when $\alpha$ is chosen to be below $\alpha_c(g_{avg})$ (here $\alpha=0.9$,
$g_i=2$, $g_f=0$ and $L=2\times 10^5$).
In both the regimes, $\mathcal{D}_n(l) \sim (\omega/n)^{1/2}$ for
$l=8,16$. (b) Fine-tuned region below $\alpha_c(g_{avg})$ where
$\mathcal{D}_n(l) \sim (\omega/n)^{3/2}$ (dark region) [even though
$\mathcal{D}_n \sim (\omega/n)^{1/2}$ both for high and low frequencies] caused
by $N_s$ changing from $2$ to $0$ due to the coalescing of two
stationary points in $0 < k < \pi$ for a certain interval in $\omega$ (
$g_f=0$ and $\alpha=1.06$ in (b)).
}%
   \label{fig6}
\end{figure}

For $\alpha < \alpha_c(g_{avg})$, the situation is  qualitatively
different. Here $N_s \neq 0$ even when $\omega \rightarrow \infty$
and hence $\mathcal{D}_n(l) \sim (\omega/n)^{1/2}$ both for fast and
slow drives. There is thus no generic reason for a dynamical phase
transition to occur as the frequency $\omega$ is varied when $\alpha
< \alpha_c(g_{avg})$, except in the fine-tuned case where $N_s$
changes from $2$ to $0$ and then back to $2$ as the frequency is
varied. Calculations using the square pulse protocol below
$\alpha_c(g_{avg})$ indeed shows that to be the case. We show the
behavior of $\mathcal{D}_n(l)$ for such a case both for fast and
slow driving frequencies in Fig.~\ref{fig6}(a) from which it is
evident that $\mathcal{D}_n(l) \sim (\omega/n)^{1/2}$ in both the
regimes of high and low frequencies. We also show an instance where
a dynamical phase transition occurs below $\alpha_c(g_{avg})$ in
Fig.~\ref{fig6}(b) when the relaxation is $\mathcal{D}_n \sim
(\omega/n)^{1/2}$ both when $\omega \rightarrow \infty$ and $\omega
\rightarrow 0$ due to the fine-tuned case of $N_s$ changing from $2$
to $0$ caused by the coalescing of two stationary points in $k \in
(0,\pi)$ in some finite-$\omega$ interval.

For the case when $\alpha > \alpha_c(g_{avg})$, as $\omega$ is
decreased further below $\omega_c^{(1)}$, the change in $N_s$ can be
non-monotonic in nature when $N_s$ is small. It is then possible
that in some frequency range, $N_s$ may revert back to zero leading
to a re-entrant behavior~\cite{SenNS2016} of the dynamical phases as
a function of $\omega$. Such re-entrance is however ruled out when
$\omega \rightarrow 0$ since $N_s \gg 1$ is this limit
(Fig.~\ref{fig3}). Due to this re-entrance effect, the phase diagram
 for the two dynamical phases has a rich structure as a function of the
frequency and amplitude of the periodic drive.

\begin{figure*}[!]
\centering
        \subfigure[]{%
            \label{fig:first}
            \includegraphics[width=0.50\textwidth]{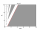}
        }%
        \subfigure[]{%
           \label{fig:second}
           \includegraphics[width=0.50\textwidth]{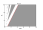}
        }\\ 
        \subfigure[]{%
            \label{fig:third}
            \includegraphics[width=0.50\textwidth]{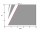}
        }%
        \subfigure[]{%
            \label{fig:fourth}
            \includegraphics[width=0.50\textwidth]{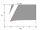}
        }%
    \caption{%
 The dynamical phase diagram calculated for the
long-ranged generalized Kitaev chain
defined in Eq.~\ref{Hamiltonian} where
$g(t)$ is driven according to the
square pulse protocol (Eq.~\ref{squarepulse})
with $g_f=0$, varying $g_i$ and
$\omega$ for different values of (a) $\alpha=4.5$, (b) $\alpha=2.5$, (c) $\alpha=1.5$ and (d) $\alpha=1.06$.
The dark (light)
regions in the phase diagrams indicate a
relaxation behavior of $\mathcal{D}_n(l) \sim (\omega/n)^{3/2}$
($\mathcal{D}_n(l) \sim (\omega/n)^{1/2}$) for any $l \ll L$.
The dotted line displayed in each
panel is obtained from the solution of
$\omega_c^{(1)}$ using Eq.~\ref{condition}.
     }%
   \label{fig7}
\end{figure*}

We illustrate the phase diagram for the dynamical phases in
Fig.~\ref{fig7} for the square pulse protocol with a fixed $g_f=0$
and varying $g_i$ and $\omega$ at different values of $\alpha$.
Firstly, for $\alpha = 4.5$ (Fig.~\ref{fig7}(a)), the phase diagram
for the dynamical phases is {\it practically indistinguishable} from
the case of $\alpha \rightarrow \infty$ where the pairing terms are
restricted to be between nearest neighbors only. Even when $\alpha
=2.5$ (Fig.~\ref{fig7}(b)), the broad features of the phase diagram
remain the same though there are now clear deviations compared to
the larger value of $\alpha$, especially in the region $\omega/\pi
\in [0,1]$. For $\alpha =1.5$ (Fig.~\ref{fig7}(c)), we first
encounter the effect that for a given amplitude $g_i$, the dynamical
phase where $\mathcal{D}_n(l) \sim (\omega/n)^{3/2}$ is completely
absent upon tuning the value of $\omega$. Furthermore, the
re-entrant region of $\mathcal{D}_n(l) \sim (\omega/n)^{3/2}$ in
$\omega/\pi \in [0,1]$ present for both $\alpha=4.5$ and $\alpha=2.5$
is completely absent. The case $\alpha=1.06$
(Fig.~\ref{fig7}(d)) shows even stronger departures compared to the
case of $\alpha \rightarrow \infty$ especially when $g_i \in [1,4]$.
When $\alpha$ is reduced further, e.g. to $\alpha=1.05$, only the
dynamical phase characterized by $\mathcal{D}_n(l) \sim
(\omega/n)^{1/2}$ survives for the shown parameter range of
$(\omega,g_i)$. This {\it discontinuous} change in the nature of the
dynamical phase diagram is because $\alpha=1.05$ is below
$\alpha_c(g_{avg})$ for the parameters $(g_i,g_f)$ considered in
Fig.~\ref{fig7}. We see that $\mathcal{D}_n(l) \sim
(\omega/n)^{1/2}$ whenever $\omega \rightarrow 0$ irrespective of
the value of $g_i$ and $\alpha$ and the complexity of the phase
diagram (Fig.~\ref{fig7}) which is consistent with $N_s \rightarrow
\infty$ as $\omega \rightarrow 0$ irrespective of the value of
$\alpha$. Finally, we also show the perfect agreement of the
location of the last dynamical transition in frequency,
$\omega_c^{(1)}$, obtained from Eq.~\ref{condition} in
Fig.~\ref{fig7} for all the different values of $\alpha$.

\section{Propagation of Mutual Information}
\label{MI}

\begin{figure}[H]
\centering {\includegraphics[width=\hsize]{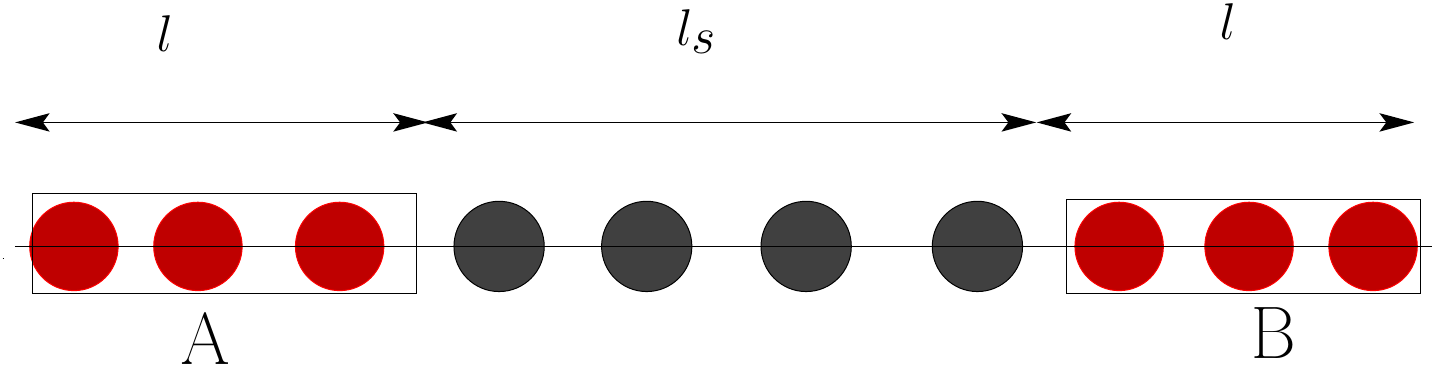}}
\caption{Schematic representation of the subsystems
$A$ and $B$, each of length $l$ and separated by a distance $l_s$,
between which the mutual information after $n$ drive cycles [denoted
by $\mathcal{I}_n(l,l_s)$] is computed.}
\label{fig8}
\end{figure}

In this section, we study the spread of entanglement
in the system described by Eq.~\ref{Hamiltonian} as a
function of space and time when $g(t)$ is a periodic function in
time. For this purpose, we monitor the mutual information
$\mathcal{I}_n(A,B)$ between two disjoint spatial regions $A$ and
$B$ to measure the total amount of correlations present between $A$
and $B$~\cite{WolfVHC2008}. $\mathcal{I}_n(A,B)$ is defined in the
following manner:
\begin{eqnarray}
\mathcal{I}_n(A,B) = S_n(A) +S_n(B)-S_n(A \cup B).
\label{MIdef}
\end{eqnarray}
For this study, we take both the regions $A$ and $B$ to contain $l$
adjacent sites each with $l_s$ sites separating these
non-overlapping regions (shown schematically in Fig. \ref{fig8}), $A
\cup B$ represents the $2l$ sites of these two subsystems together,
and $S_n(R)$ is the entanglement entropy of the subsystem $R$ after
$n$ drive cycles using Eq.~\ref{Svn}. Henceforth, we will denote the
mutual information between two disjoint subsystems by
$\mathcal{I}_n(l,l_s)$. $\mathcal{I}_n(l,l_s)$ has the property that
it is positive and can only vanish if $\rho_r(A \cup B) = \rho_r(A)
\otimes \rho_r(B)$. Therefore, starting from an unentangled state at
$n=0$, $\mathcal{I}_n(l,l_s)$ provides an unbiased measure of when
the two regions $A$ and $B$ get entangled with each other as $n$ is
progressively increased.

\begin{figure*}[!]
\centering
        \subfigure[]{%
            \label{fig:first}
            \includegraphics[width=0.50\textwidth]{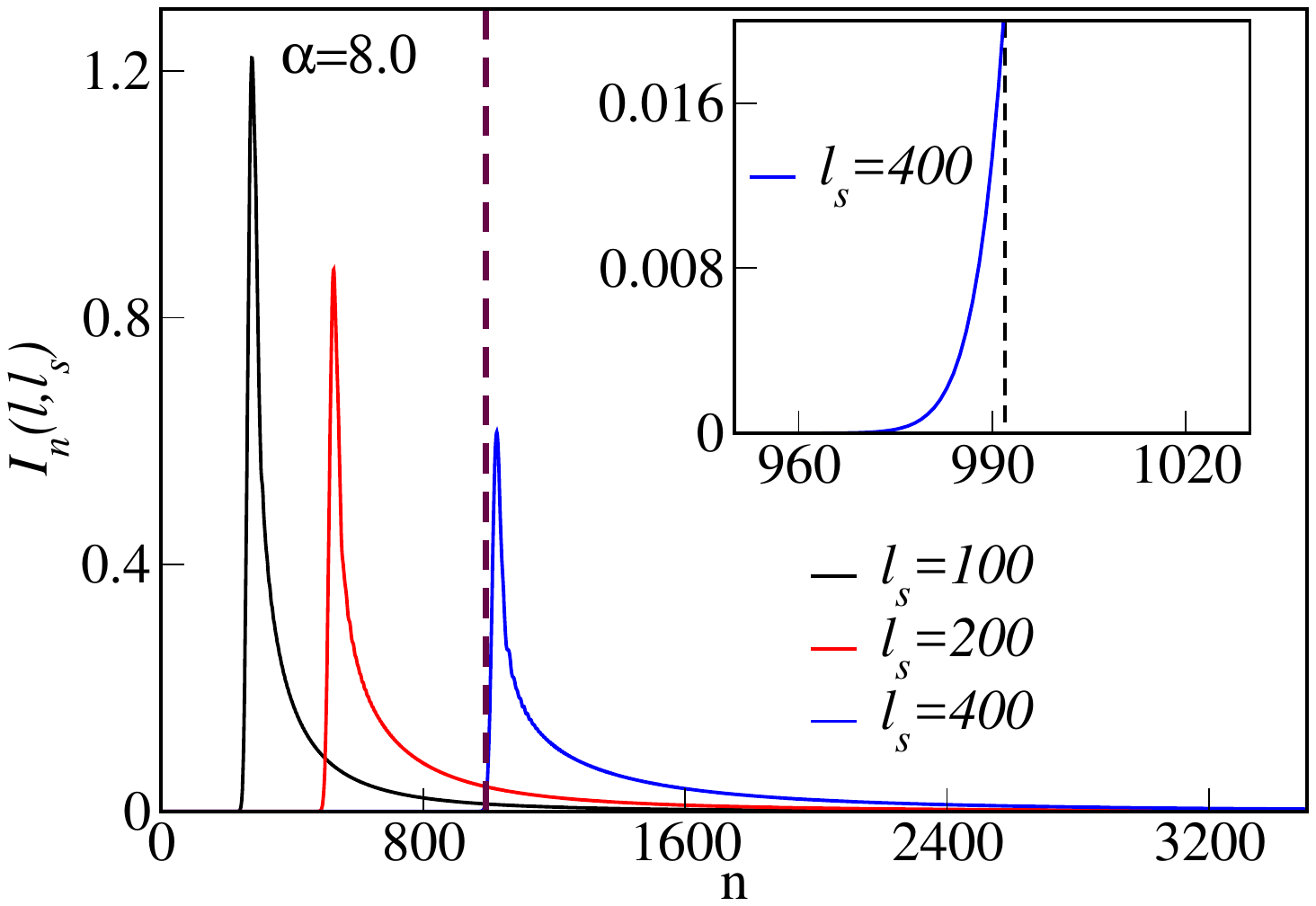}
        }%
        \subfigure[]{%
           \label{fig:second}
           \includegraphics[width=0.50\textwidth]{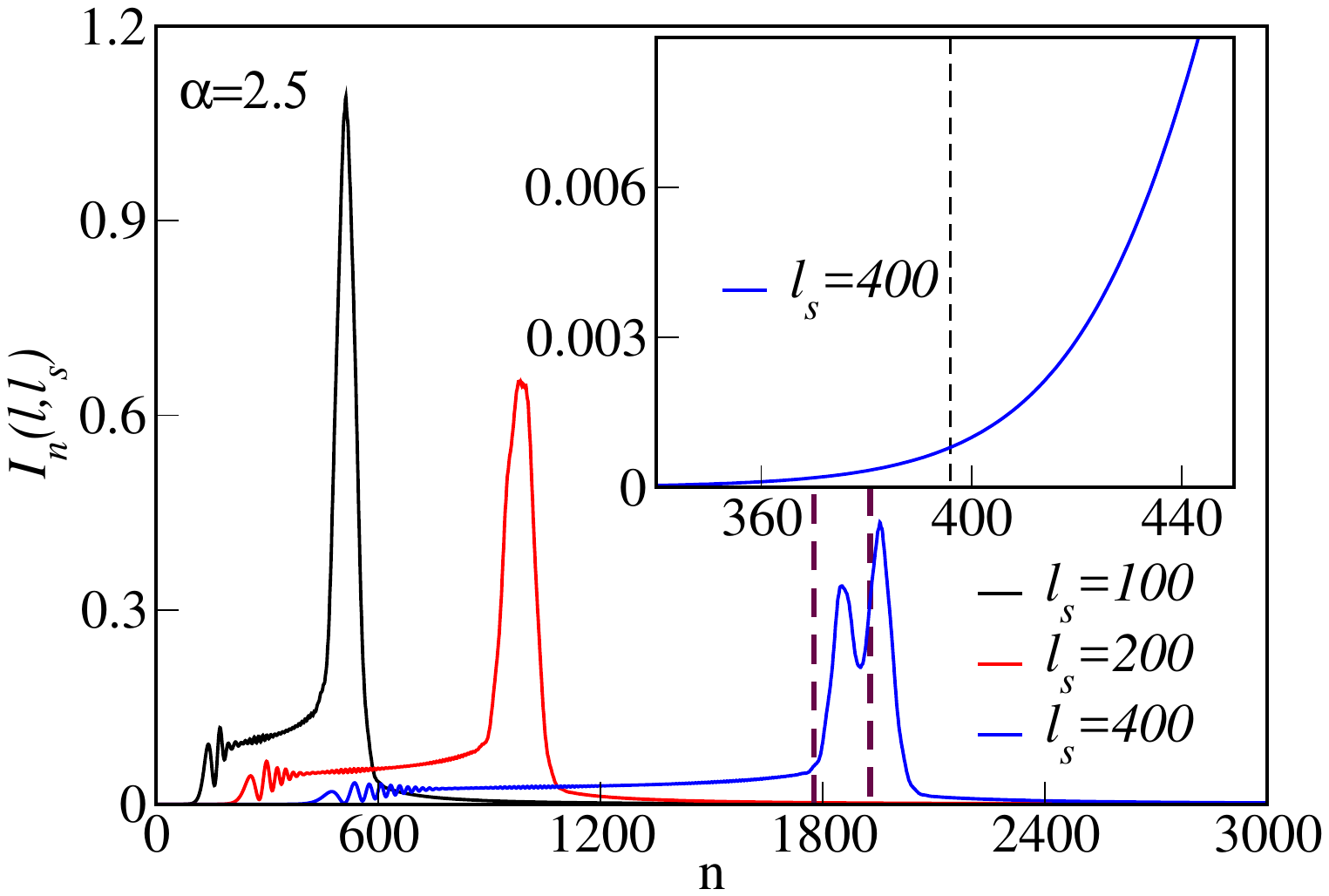}
        }\\ 
        \subfigure[]{%
            \label{fig:third}
            \includegraphics[width=0.50\textwidth]{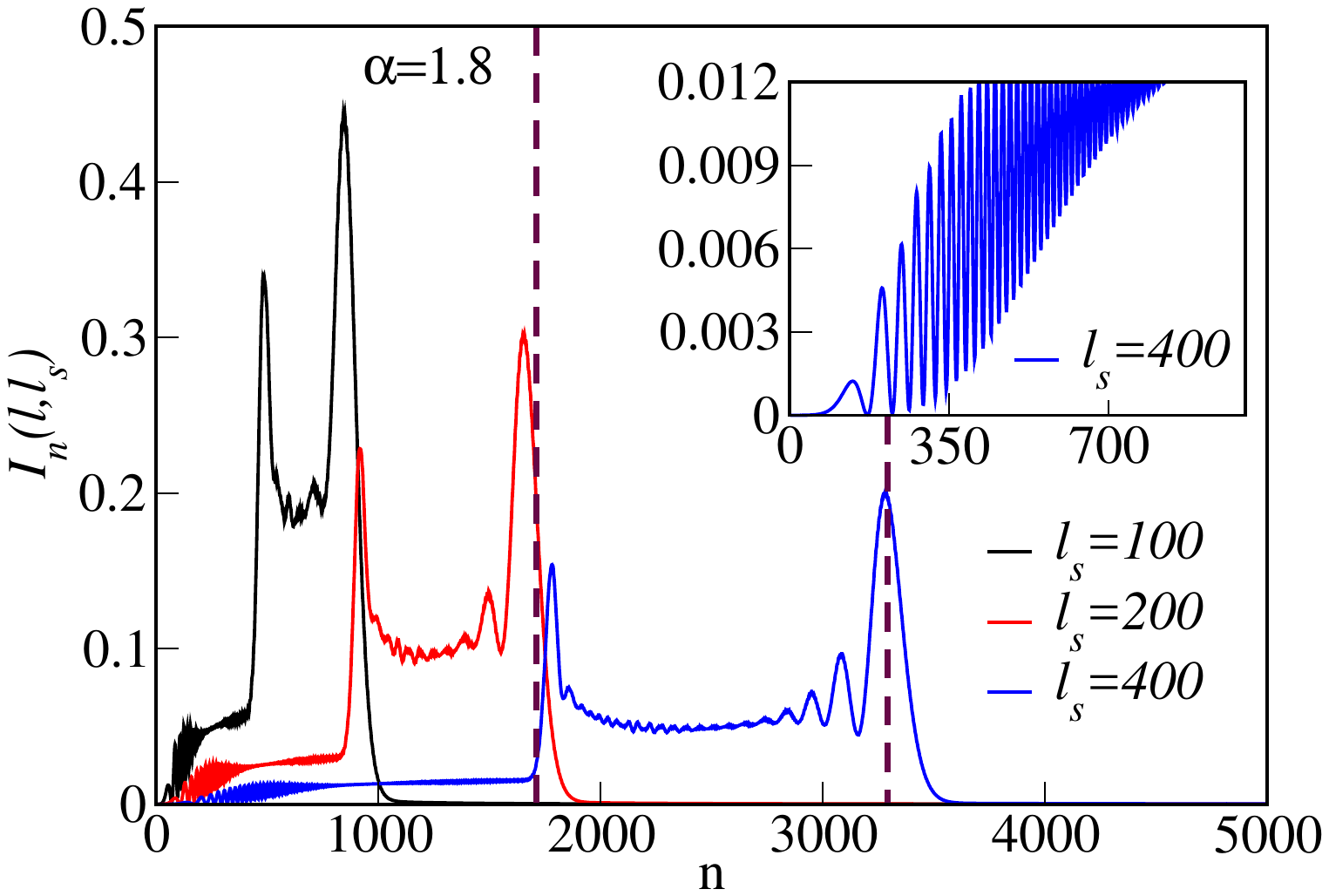}
        }%
        \subfigure[]{%
            \label{fig:fourth}
            \includegraphics[width=0.50\textwidth]{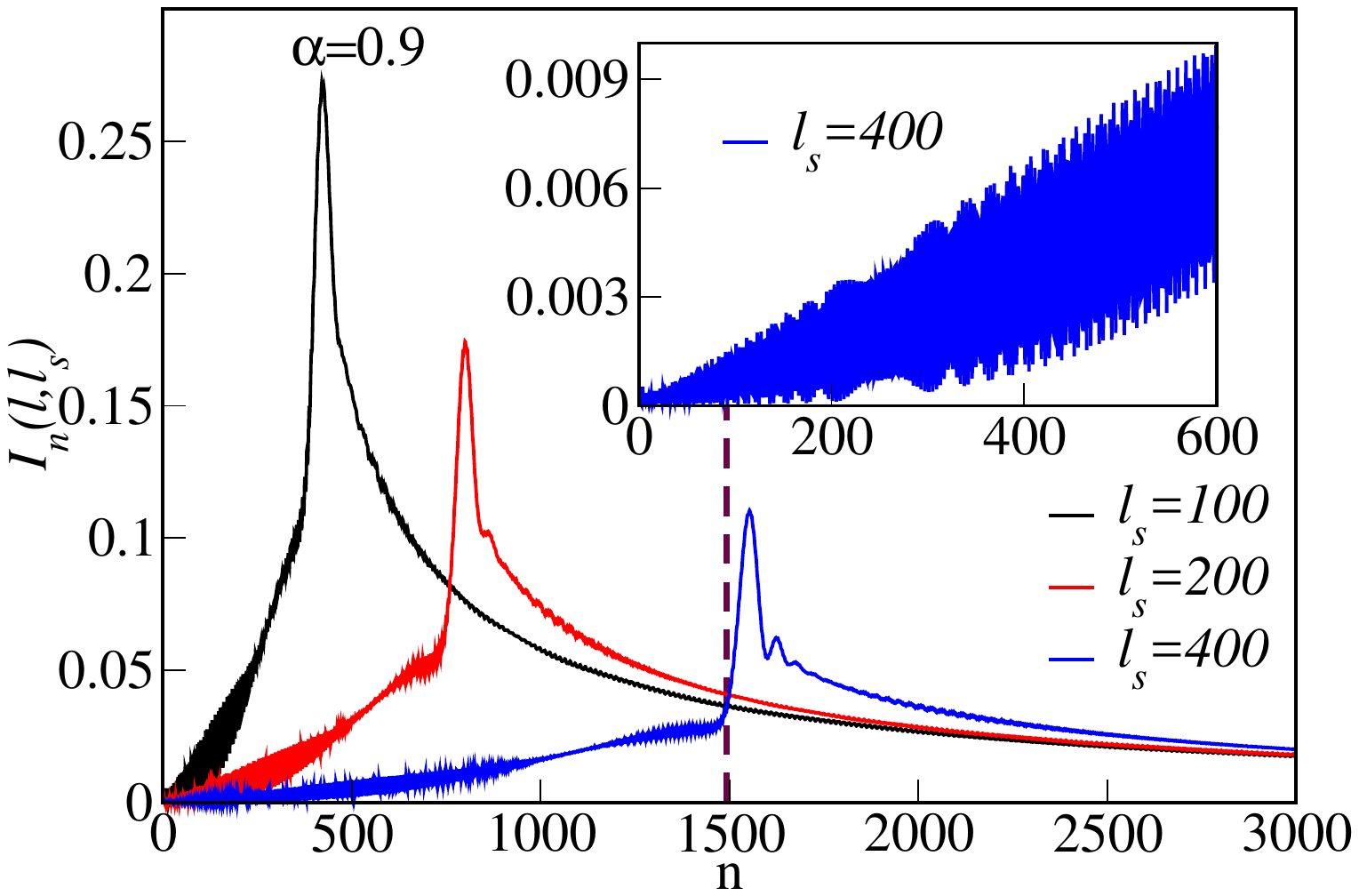}
        }%
    \caption{%
The propagation of mutual information $\mathcal{I}_n(l,l_s)$ where
$l=10$ and $l_s=100$ (black), $l_s=200$ (red), $l_s=400$ (blue).
The periodic drive
parameters are $g_i=2$, $g_f=0$, and $\omega/\pi=10$ with system
size $L =2\times 10^5$. The state at $n=0$ is the vacuum state of
the fermions. The four panels show data for (a) $\alpha=8.0$, (b)
$\alpha=2.5$, (c) $\alpha=1.8$ and (d) $\alpha=0.9$.
The insets of the panels in (a) and (b) (data for $l_s=400$) show
that mutual information becomes non-zero only after a finite $n$.
Dotted vertical lines in the insets are at
$n=l_s/(2T(v_g^F)^{\mathrm{max}})$ with $l_s=400$.
The insets of panels in (c) and (d) (data for $l_s=400$)
show that mutual information
becomes non-zero immediately for any $n>0$.
 The dotted lines
in the main panels of (a),(b),(c), and (d)
are at $n=l_s/(2T |v_g^F(k^*)|)$ (with $l_s=400$ here) such that $D_g^F(k
\rightarrow k^*) \rightarrow \infty$.
     }%
   \label{fig9}
\end{figure*}

The behavior of $\mathcal{I}_n(l,l_s)$ is shown in Fig.~\ref{fig9}
for the power-law decay exponent $\alpha=8.0$ (Fig.~\ref{fig9}(a)),
$\alpha=2.5$ (Fig.~\ref{fig9}(b)), $\alpha=1.8$ (Fig.~\ref{fig9}(c))
and $\alpha=0.9$ (Fig.~\ref{fig9}(d)) respectively for a square
pulse protocol (Eq.~\ref{squarepulse}) with the parameters being
$g_i=2$, $g_f=0$, and $\omega/\pi=10$. The pure state at $n=0$ is
the vacuum state of the fermions. We take a fixed size of $l=10$
adjacent sites for both the regions $A$ and $B$ and show the results
for $\mathcal{I}_n(l,l_s)$ for a separation of $l_s=100, 200$ and
$400$ sites as a function of the stroboscopic time $n$ in
Fig.~\ref{fig9}. For $\alpha=8.0$ (Fig.~\ref{fig9}(a)), we have
checked that the behavior of the mutual information is practically
indistinguishable from the short-ranged case where the pairing terms
are restricted to be between nearest neighbors (i.e., $\alpha
\rightarrow \infty$). $\mathcal{I}_n(l,l_s)$ becomes non-zero only
after a finite $n$, the value of which increases linearly with the
distance between the disjoint blocks ($l_s$) (Fig.~\ref{fig9}(a)),
thus clearly showing the light cone effect with a well-defined
velocity. For a fixed $l_s$, $\mathcal{I}_n(l,l_s)$ shows a strong
peak at a value of $n$ close to where it first becomes non-zero
(inset of Fig.~\ref{fig9}(a)). For $\alpha=2.5$
(Fig.~\ref{fig9}(b)), there are already significant deviations
compared to $\alpha \rightarrow \infty$. For example, the peak in
$\mathcal{I}_n(l,l_s)$ for a fixed $l_s$ as a function of $n$ {\it{does
not}} appear soon after it first turns non-zero (inset of
Fig.~\ref{fig9}(b)) but only at a much later value of $n$ unlike
when $\alpha=8.0$. However, the mutual information again first turns
non-zero only after a finite $n$ that scales linearly with the
distance between the blocks $l_s$. Moreover, the position of the
peak in the mutual information that emerges only at a much later $n$
also scales linearly with increasing $l_s$ with a {\it different}
velocity that is distinct from the light cone velocity. In
Fig.~\ref{fig9}(c),(d), we display the effect of lowering $\alpha$
further on the propagation of mutual information. Both for
$\alpha=1.8$ (Fig.~\ref{fig9}(c)) and for $\alpha=0.9$
(Fig.~\ref{fig9}(d)), the mutual information behaves completely
differently from the cases shown in Fig.~\ref{fig9}(a),(b) in that
no matter how large the separation between the blocks ($l_s$), the
mutual information is always non-zero for any $n>0$ which implies
that the blocks become entangled with each other {\it
instantaneously} showing the absence of a strict light cone effect.
The immediate growth of the mutual information for any $n>0$ is
demonstrated more clearly in the insets of the corresponding figures
in Fig.\ref{fig9}(c),(d). However, in spite of the absence of a
light cone effect, there are still clear features in terms of local
peaks of the mutual information as a function of $n$ where the peak
positions in $n$ increase linearly with $l_s$ (main panels of
Fig.~\ref{fig9}(c),(d)). This means that one can associate the
notion of a well-defined velocity for such features even at small
$\alpha$ where there is an instantaneous propagation of the
entanglement.

The results displayed in Fig.~\ref{fig9}
for $\omega/\pi=10$ can be qualitatively understood by using results
from previous studies of quantum quenches in such long-ranged
models. We note that at large $\omega$, the Floquet Hamiltonian that
describes the stroboscopic time evolution equals the time-averaged
Hamiltonian over one drive cycle $\bar{H}$ as $\omega \rightarrow
\infty$ and the problem can be formally mapped to a global quantum
quench with the post-quench Hamiltonian being equal to $\bar{H}$. We
can then directly apply the results obtained in
Refs.~\onlinecite{HaukeT2013, RegemortelSW2016,BuyskikhFSED2016}
which we summarize below. The group velocity of the quasiparticles
at momentum $k$ can be obtained from $v_g(k)=d|\vec{\epsilon}_k|/dk$
where $|\vec{\epsilon}_k|$ is given in Eq.~\ref{largew} when $\omega
\rightarrow \infty$. The maximum of the magnitude of the group
velocity $v_g(k)$ as a function of $k$, which we denote by
$v_g^{\mathrm{max}}$, is finite~\cite{HaukeT2013, RegemortelSW2016,
BuyskikhFSED2016} when $\alpha > 2$, which justifies the presence of
the light cone effect for global quenches even in such long-ranged
systems. However, $v_g^{\mathrm{max}} \rightarrow \infty$ when
$\alpha \rightarrow 2^+$. Near $k=0$, the dispersion relation of the
quasiparticle energy behaves as~\cite{HaukeT2013,
RegemortelSW2016,BuyskikhFSED2016}
\begin{eqnarray}
|\vec{\epsilon}_k|_{\omega \rightarrow \infty} \sim \epsilon_0 + A k^{\alpha-1}.
\label{quenchdispersion}
\end{eqnarray}
Thus the group velocity near $k=0$ diverges as $k^{\alpha-2}$ for
any $\alpha < 2$. The spectrum is also unbounded as $k \rightarrow
0$ when $\alpha <  1$. Thus, there is no sharp light cone for a
quantum quench when $\alpha < 2$, consistent with the behavior
displayed in Fig.~\ref{fig9}(c),(d) for a large driving frequency.
\begin{figure}
\begin{center}
        \subfigure[]{%
            \label{fig:first}
            \includegraphics[width=0.50\textwidth]{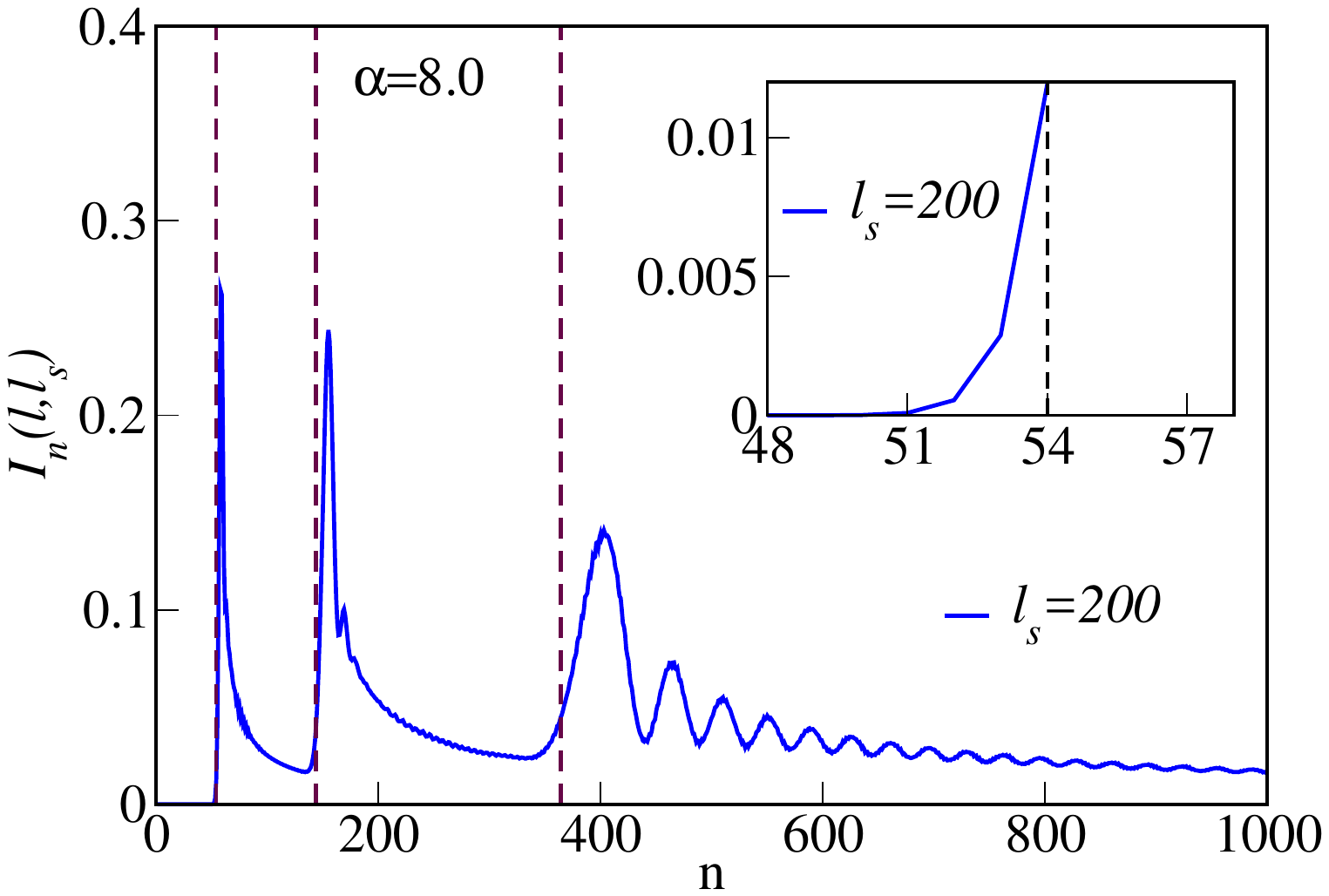}
        }\\%
        \subfigure[]{%
            \label{fig:third}
            \includegraphics[width=0.50\textwidth]{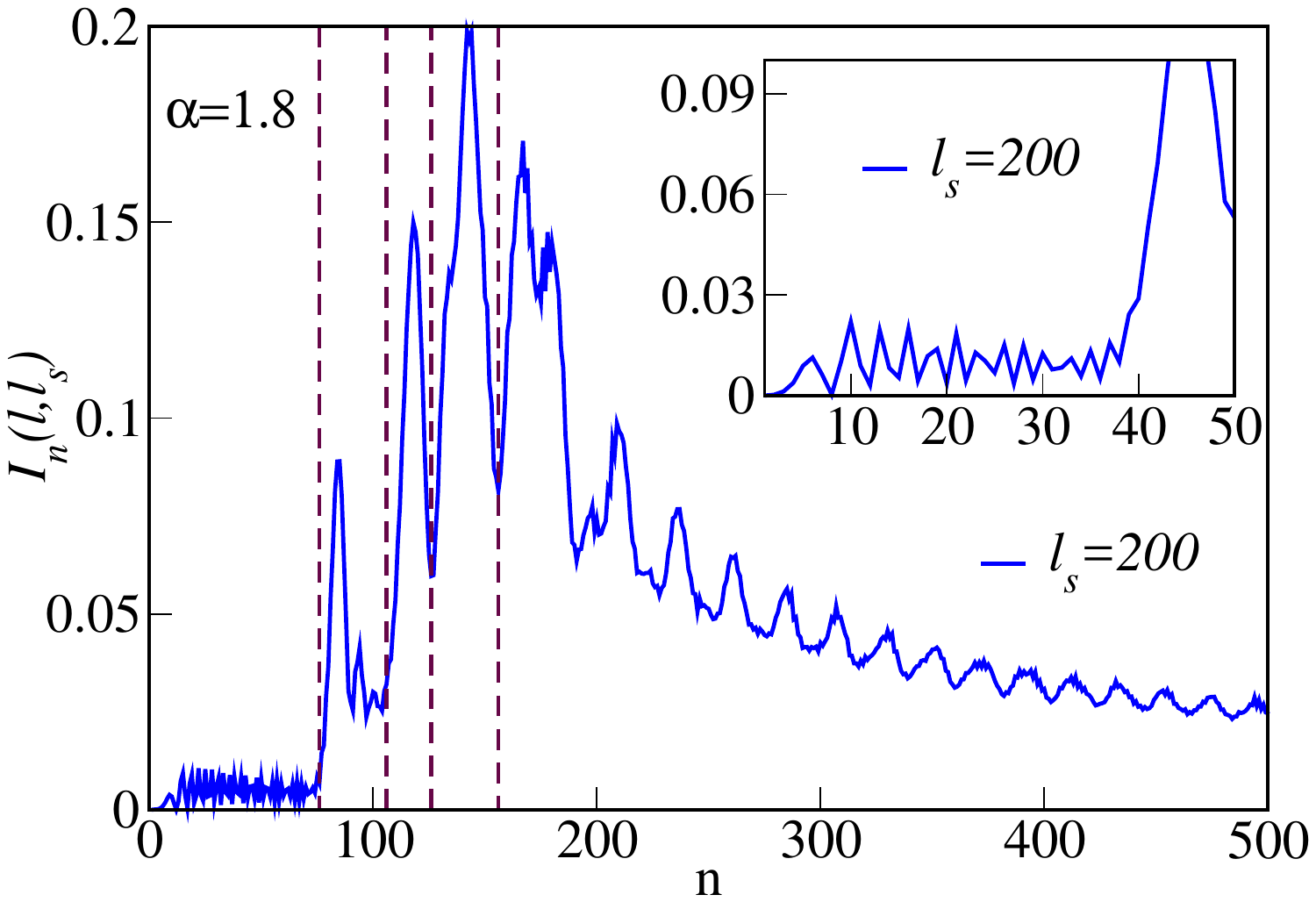}
        }%
    \end{center}
    \caption{%
The propagation of mutual information $\mathcal{I}_n(l,l_s)$
where $l=10$ and $l_s=200$.
The periodic drive parameters are $g_i=2$, $g_f=0$,
and $\omega/\pi=0.5$ with system size $L =2\times 10^5$.
The state at $n=0$
is the vacuum state of the fermions.
The two panels show data for (a) $\alpha=8.0$ and (b) $\alpha=1.8$.
The insets of these panels (data for $l_s=200$) show that in
(a) mutual information becomes non-zero only after a finite $n$.
Dotted vertical line in the inset is at $n=l_s/(2T(v_g^F)^{\mathrm{max}})$
with $l_s=200$; and in (b) mutual information
becomes non-zero immediately for any $n>0$.
The dotted lines in the main panels are at
$n=l_s/(2T |v_g^F(k^*)|)$ (with $l_s=200$ here) such that $D_g^F(k
\rightarrow k^*) \rightarrow \infty$.
     }%
   \label{fig10}
\end{figure}

At any finite $\omega$, the spreading of the mutual information
deviates from the global quantum quench. Then, a natural question
that arises is that when do qualitatively new features appear in the
entanglement propagation as the driving frequency of the periodic
protocol is decreased? In Fig.~\ref{fig10}, we show the mutual
information propagation for the same combination of $\alpha,g_i$ and
$g_f$ as in Fig.~\ref{fig9}(a) and Fig.~\ref{fig9}(c) but at a lower
driving frequency of $\omega/\pi=0.5$. The mutual information
profile is now completely different compared to the case where
$\omega/\pi=10$ (which was similar to that of a global quench) and
has much more structure. Crucially, there is still a well-defined
light cone effect for $\alpha>2$ (as shown for $\alpha=8.0$ in
Fig.~\ref{fig10}(a), inset) while the entanglement builds up
immediately when $\alpha < 2$ (as shown for $\alpha =1.8$ in
Fig.~\ref{fig10}(b), inset) even at low $\omega$. In particular, for
large $\alpha$, the space-time propagation of the mutual information
shows a simple behavior with a single sharp light cone front when
the driving protocol frequency is large (Fig.~\ref{fig9}(a)), but
clear {\it multiple} light cone fronts with distinct velocities for
lower $\omega$ (as can be seen in Fig.~\ref{fig10}(a)).

\begin{figure*}[!]
\centering
        \subfigure[]{
            \label{fig:first}
            \includegraphics[width=0.40\textwidth]{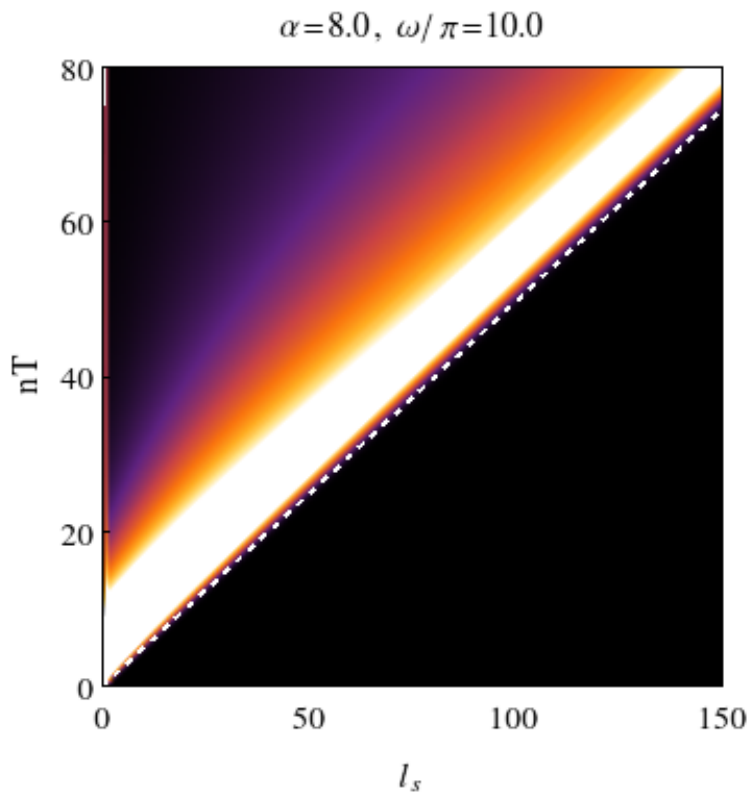}
        }
        \subfigure[]{
           \label{fig:second}
           \includegraphics[width=0.40\textwidth]{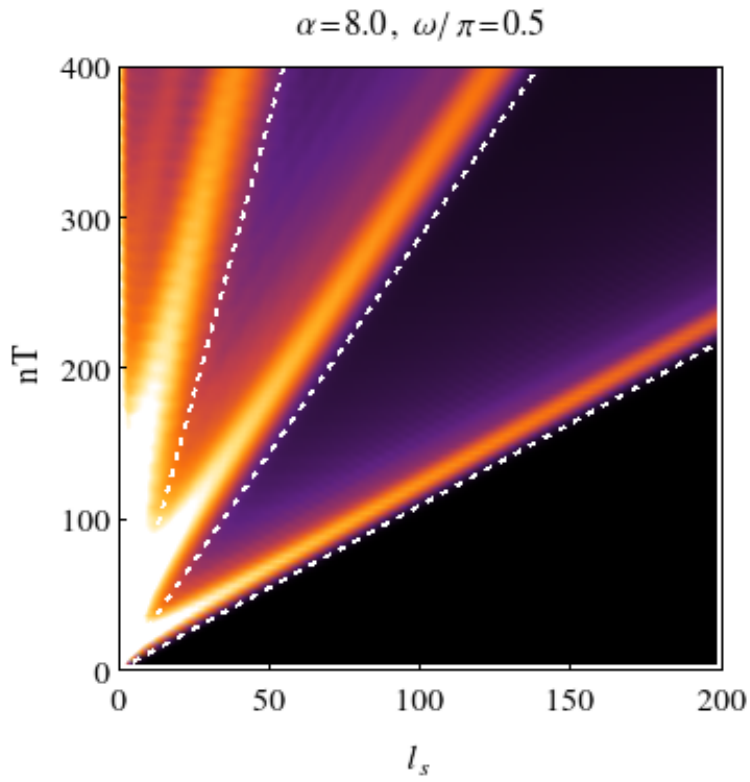}
        }\\ 
        \subfigure[]{%
            \label{fig:third}
            \includegraphics[width=0.40\textwidth]{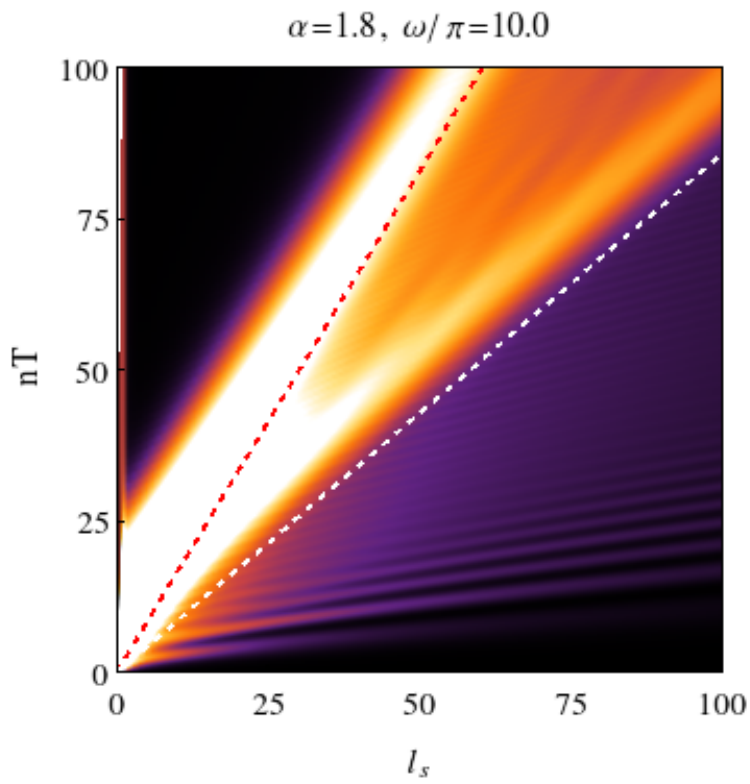}
        }%
        \subfigure[]{%
            \label{fig:fourth}
            \includegraphics[width=0.40\textwidth]{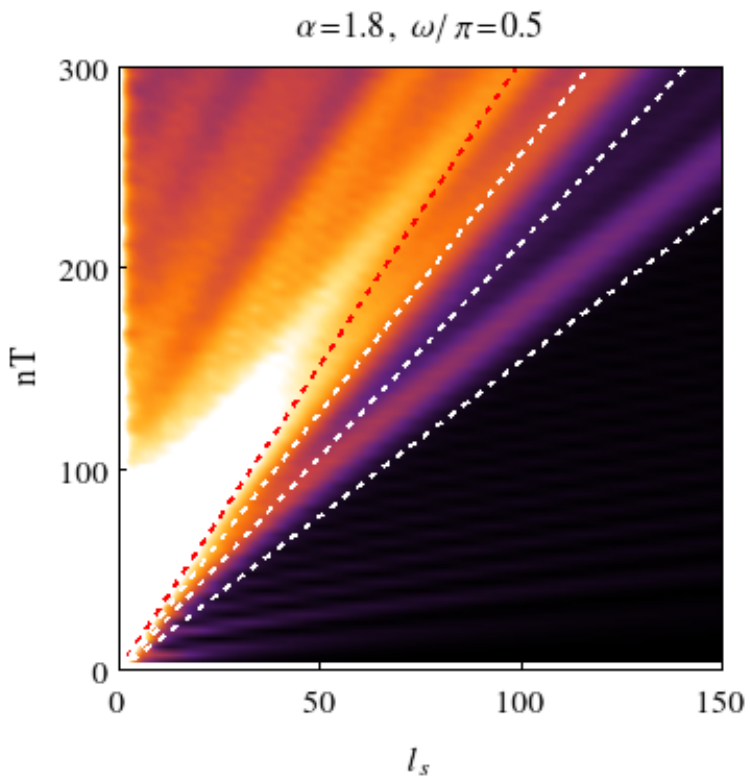}
        }

    \caption{ The propagation of mutual information $\mathcal{I}_n(l,l_s)$ shown
as a function of the subsystem separation ($l_s$) and stroboscopic
time ($nT$) where each subsystem has $l=10$ sites. The state at
$n=0$ is the vacuum state of the fermions. The periodic drive
parameters are $g_i=2$, $g_f=0$, and (a) $\alpha=8.0$,
$\omega/\pi=10.0$ (b) $\alpha=8.0$, $\omega/\pi=0.5$ (c)
$\alpha=1.8$, $\omega/\pi=10.0$ (d) $\alpha=1.8$, $\omega/\pi=0.5$.
Here bright (dark) colour represents a higher (lower) value of
mutual information. The dotted lines displayed in panels (a), (b),
(c), (d) have slopes equal to $1/(2|v_g^F(k^*)|)$ such that $D_g^F(k
\rightarrow k^*) \rightarrow \infty$.
     }
   \label{fig11}
\end{figure*}

The presence (absence) of light cone like features in the spreading
of mutual information in space-time for $\alpha > 2$ ($\alpha <2$)
at any drive frequency $\omega$ can be easily seen by plotting
$\mathcal{I}_n(l,l_s)$ as a function of both the subsystem
separation ($l_s$) and the stroboscopic time ($nT$) as shown in
Fig.~\ref{fig11}. For $\alpha =8.0$, we see a single light cone
feature for a large drive frequency $\omega/\pi=10.0$
(Fig.~\ref{fig11}(a)). For the same $\alpha$, we see the presence of
multiple light cone features in the mutual information propagation
for a lower drive frequency of $\omega/\pi=0.5$
(Fig.~\ref{fig11}(b)). For a low $\alpha (=1.8)$, we can see that
there is no sharp light cone effect irrespective of whether the
drive frequency $\omega$ is large (Fig.~\ref{fig11}(c)) or small
(Fig.~\ref{fig11}(d)). Also, we can clearly see that the mutual
information propagation in space-time for low $\alpha$ is
qualitatively different at $\omega/\pi=0.5$ compared to the
high-frequency drive frequency case ($\omega/\pi=10.0$).

\begin{figure}
\begin{center}
       \subfigure[]{%
            \label{fig:first}
            \includegraphics[width=0.5\textwidth]{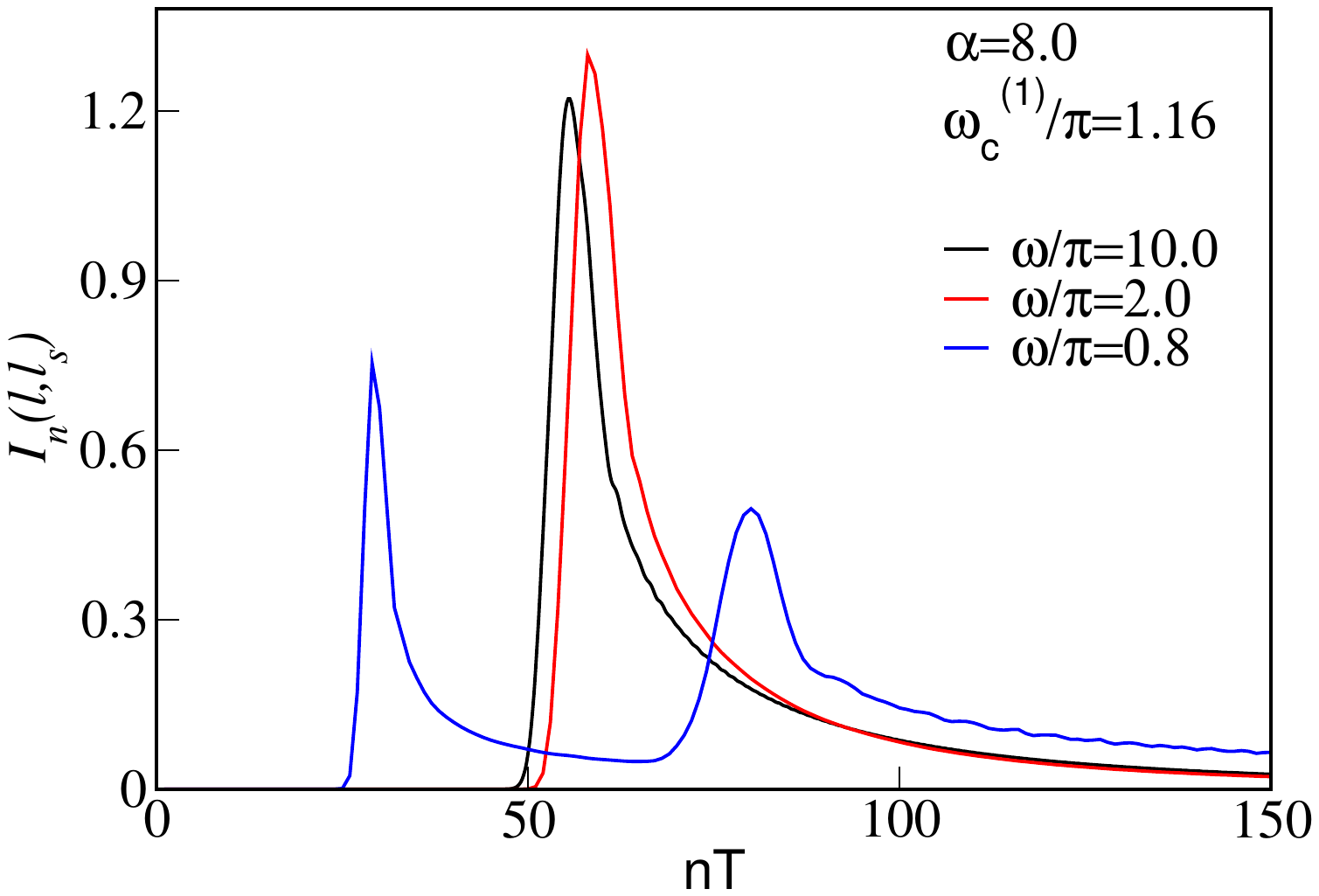}
        } \\%
        \subfigure[]{%
           \label{fig:second}
           \includegraphics[width=0.5\textwidth]{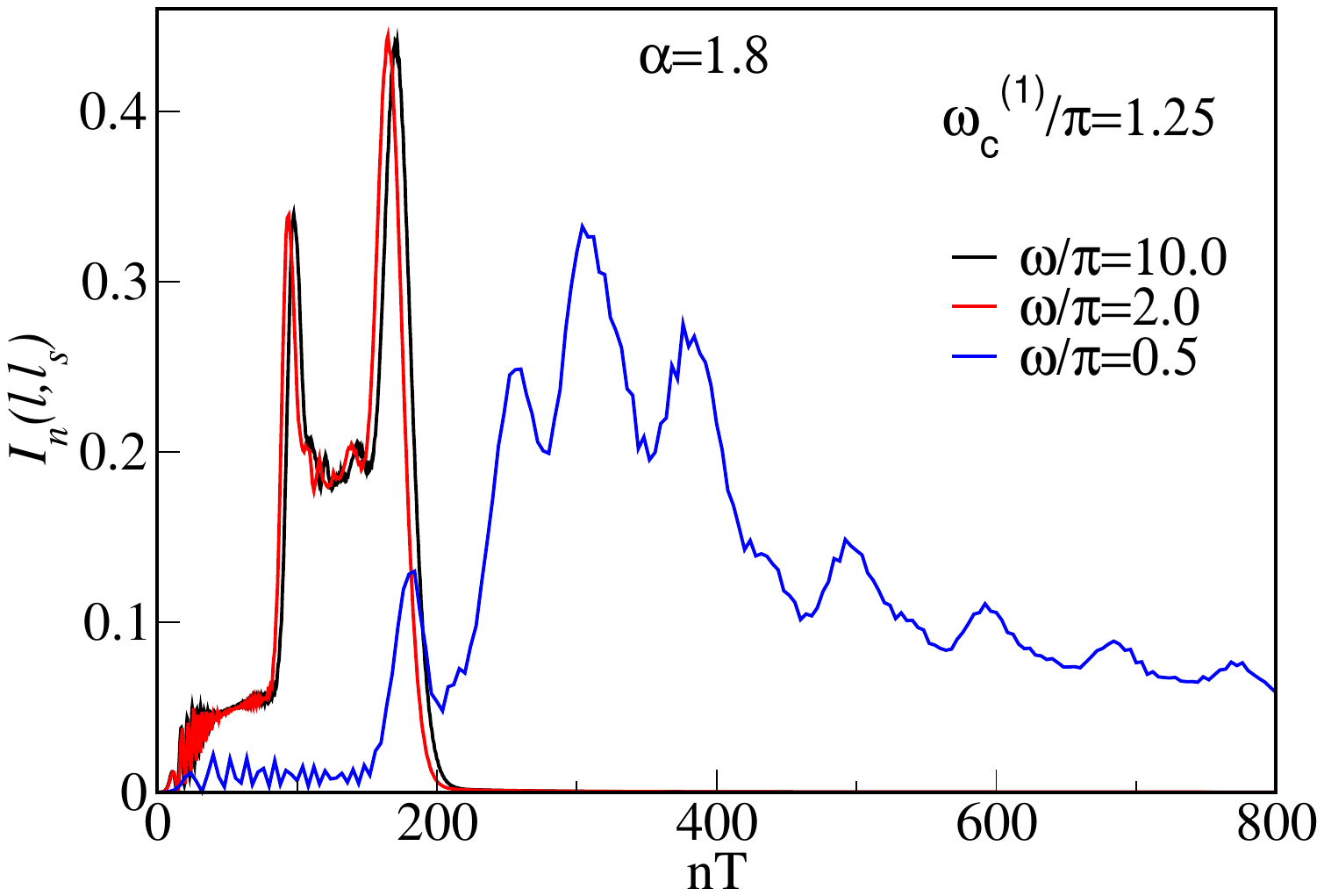}
        } 
    \end{center}
    \caption{%
We show the behavior of $\mathcal{I}_n(l,l_s)$ for $l=10$ and
$l_s=100$ with $g_i=2$ and $g_f=0$ as a function of $nT$ where $T$
is the time-period of the drive for multiple driving frequencies
with (a) $\alpha=8.0$ and (b) $\alpha=1.8$. In both panels, the
mutual information shows no new features when $\omega \in
(\omega_c^{(1)}, \infty)$ and qualitatively new features when
$\omega < \omega_c^{(1)}$ irrespective of whether there is a light
cone effect ($\alpha >2$ as in (a)) or not ($\alpha \le 2$ as in
panel (b)).
     }%
   \label{fig12}
\end{figure}

In Fig.~\ref{fig12}, we see that the appearance of new features in
the propagation of the mutual information in space-time is
intimately tied to the {\it last} dynamical phase transition in
frequency for any $\alpha > \alpha_c$ as $\omega$ is varied in the
range $[0,\infty)$ (discussed in Sec.~\ref{dynamicaltransitions}).
More precisely, for a driving frequency $\omega \in
(\omega_c^{(1)},\infty)$, the mutual information spreading shows no
new features compared to the $\omega \rightarrow \infty$ limit
irrespective of whether $\alpha
>2$ (Fig.~\ref{fig12}(a)) (where there is a
strict light cone effect present at any $\omega$) or $\alpha <2$
(Fig.~\ref{fig12}(b)) (where there is no light cone effect at any
$\omega$). When $\omega <
\omega_c^{(1)}$, qualitatively new features emerge both when $\alpha
> 2$ (Fig.~\ref{fig12}(a)) and $\alpha <2$ (Fig.~\ref{fig12}(b)).

\begin{figure}
\begin{center}
       \subfigure[]{%
            \label{fig:first}
            \includegraphics[width=0.5\textwidth]{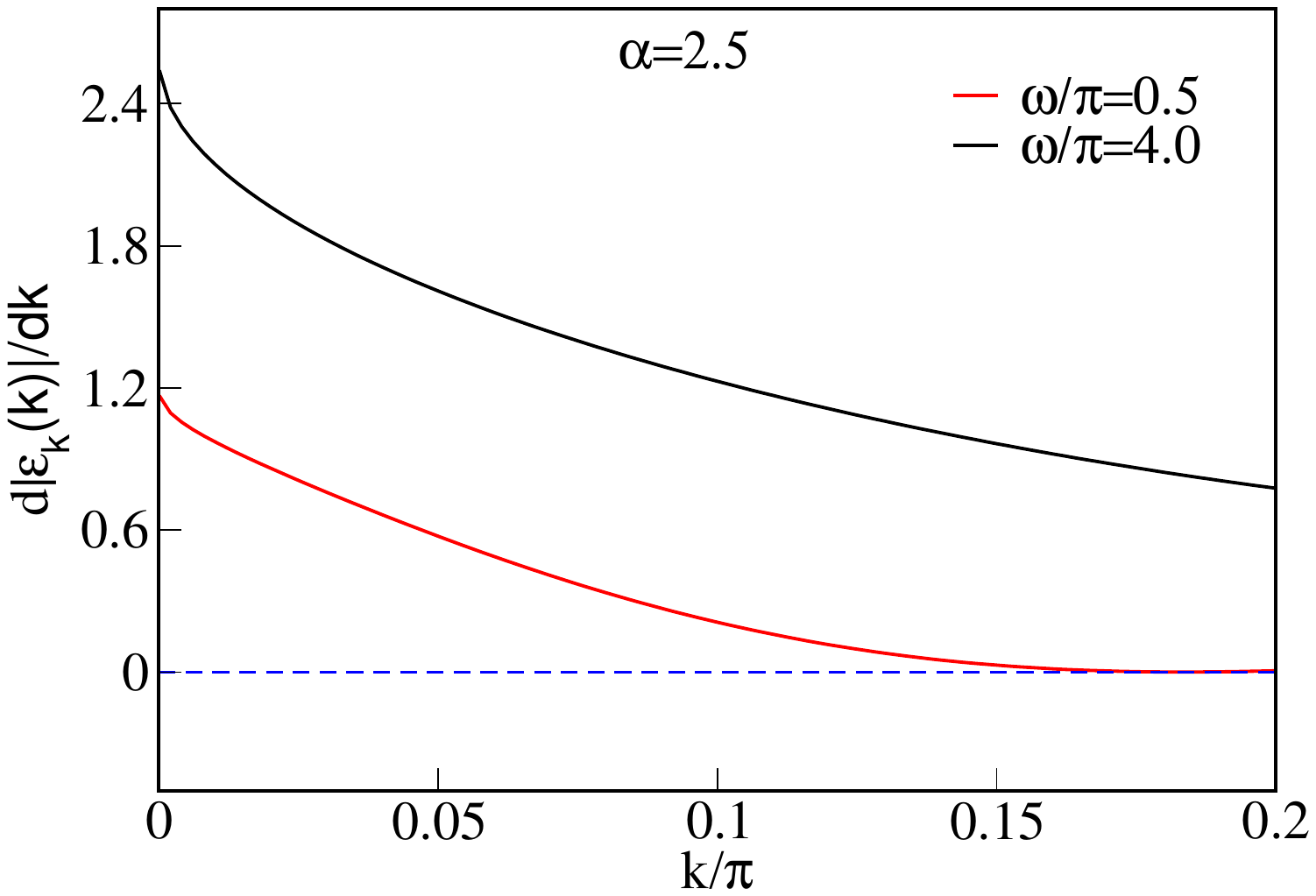}
        } \\%
        \subfigure[]{%
           \label{fig:second}
           \includegraphics[width=0.5\textwidth]{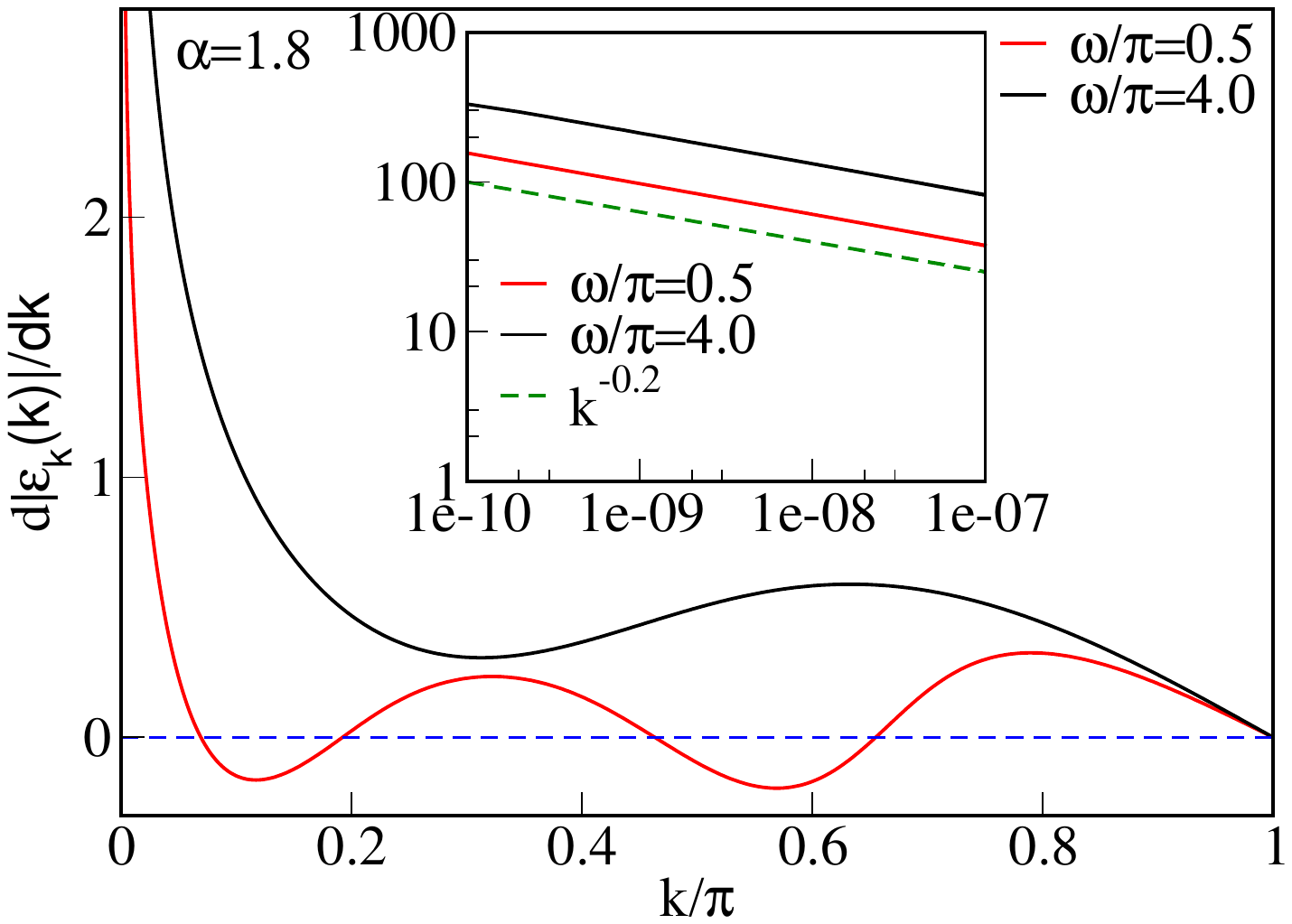}
        } 
    \end{center}
    \caption{%
 We show here that $(v_g^{F})^{\mathrm{max}}$  is finite for
(a) $\alpha > 2$ (data for $\alpha=2.5$) and diverges for
(b) $\alpha <2$ (data for $\alpha=1.8$) irrespective of the value
of the driving frequency $\omega$. Here we use
the square pulse protocol and $g_i=2.0, g_f=0.0$ for finite $\omega$. 
The inset of (b) shows that the divergence near $k=0$ 
is of the form $k^{\alpha-2}$
for $\alpha<2$ even at finite $\omega$.
     }%
   \label{fig13}
\end{figure}

To understand generic features of the spread of entanglement in
space as a function of the stroboscopic time and its dependence on
the driving frequency for the generalized Kitaev chain, it is
sufficient to look at the behavior of $\delta C_{ij}(n)$ and $\delta
F_{ij}(n)$ (Eq. \ref{cfijexp}). For brevity, we only analyze $\delta
C_{ij}(n)$ (since $\delta F_{ij}(n)$ leads to similar conclusions)
and focus on the ``space-time scaling limit''~\cite{CalabreseEF2011}
where both $l_s=(i-j) \rightarrow \infty$ and $n \rightarrow
\infty$, with $l_s/n=u_s$ fixed. Expressing the integrand in terms
of $u_s$ and $n$, we get
\begin{eqnarray}
\delta C_{ij}(n) &=& \int_0^\pi \frac{dk}{8 \pi}
(\hat{n}_{k3}^2-1 ) \left( e^{i\Phi_{+}(k)n}+e^{i\Phi_{-}(k)n}+\mathrm{c.c.} \right) \nonumber \\
\Phi_{\pm} (k) &=& (ku_s \pm 2|\vec{\epsilon}_k|T)
\label{spacetime}
\end{eqnarray}
Thus, along the line $l_s/n=u_s$, the integral in Eq.~\ref{spacetime} is
dominated by the stationary points of $\Phi_{\pm}(k)$ given by the
$k$ values (denote by $k^*$) where $d\Phi_{\pm}(k)/dk=0$ which gives
\begin{eqnarray}
2 v_g^F(k^*)T = \pm u_s
\label{spcondition}
\end{eqnarray}
where we have defined the ``Floquet group velocity'' of the
quasiparticles at momentum $k$ as $v_g^F(k) =
d|\vec{\epsilon}_k|/dk$ (here, we stress again that we are working
in the reduced zone scheme as explained below Eq.~\ref{Uform}). We
numerically see from Fig.~\ref{fig13} that the maximum magnitude of
$v_g^F(k)$ in the BZ, which we denote by $(v_g^{F})^{\mathrm{max}}$,
is finite for $\alpha > 2$ and diverges for $\alpha <2$ irrespective
of the value of $\omega$ using the square pulse protocol
(Eq.~\ref{squarepulse}), and not just when $\omega \rightarrow
\infty$ where the problem reduces to that of a global quench.
Furthermore, the divergence in $v_g^F(k)$ arises when $k \rightarrow
0$ and is of the form $k^{\alpha-2}$ for $\alpha<2$ irrespective 
of the value of $\omega$ (as shown in the inset of Fig.~\ref{fig13}
(b)). This explains the build up of the mutual information
immediately for any $n>0$ as shown in the inset of
Fig.~\ref{fig9}(c) and Fig.~\ref{fig10}(b)
when $\alpha=1.8$, unlike the case shown in
(inset of) Fig.~\ref{fig9}(a) and Fig.~\ref{fig10}(a)
where $\alpha=8.0$.

We now consider the behavior of
mutual information for a fixed $l_s$ as a function of $n$ (as shown
in Fig.~\ref{fig9} and Fig.~\ref{fig10}). At
large $n$, $\delta C_{ij}(n)$ will receive a contribution from a
stationary point $k^*$ whenever Eq.~\ref{spcondition} is satisfied
for a $k^* \in [0,\pi]$. From this, it is immediately clear that if
\begin{eqnarray}
n < n_c\left(=\frac{l_s}{2T (v_g^F)^{\mathrm{max}}} \right)
\label{nc}
\end{eqnarray}
then Eq.~\ref{spcondition} does not have any solution, and $\delta
C_{ij}(n)$ is vanishingly small. This explains the resulting light
cone effect whenever $(v_g^F)^{\mathrm{max}}$ is finite, since
otherwise $n_c \rightarrow 0$. In fact, the mutual information
decays exponentially as $\exp(-(n_c-n)/\xi(\alpha,\omega))$ for $n <
n_c$ when $l_s$ is large (insets of Figs.~\ref{fig9}(a),(b) and
Fig.~\ref{fig10}(a)), where $\xi(\alpha,\omega) \rightarrow \infty$
as $\alpha \rightarrow 2^+$ since $(v_g^F)^{\mathrm{max}}$ diverges
below $\alpha=2$ (Fig.~\ref{fig13}) for any $\omega$.


For $n>n_c$, there may be a solution at some $k^*$ where
Eq.~\ref{spcondition} is satisfied at a particular $n$. Apart from
an oscillatory sinusoidal factor, this contribution from $k^*$ will
scale as (the form of the stationary point contribution may be read
off from Eq.~\ref{stationarypt})
\begin{eqnarray}
\sqrt{\frac{D_g^F(k^*)}{nT}}, \mbox{~~} \mathrm{with} \mbox{~~}D_g^F(k)=\frac{1}{\pi}|dv_g^F(k)/dk|^{-1},
\label{ampsp}
\end{eqnarray}
where $D_g^F(k)$ can be interpreted as a density of states in
velocity as a function of $k$ since it can be written as $D_g^F(k) =
(1/\pi) \int_0^\pi dk \delta (v-v_g^F(k))$. Thus at a fixed $l_s$,
mutual information will then show strong features in the
neighborhood of $n=l_s/(2T|v_g^F(k^*)|)$ (the stationary point
condition of Eq.~\ref{stationarypt}) when $D_g^F(k^*) \rightarrow
\infty$. In Fig.~\ref{fig9}(a),(b),(c),(d) and
Fig.~\ref{fig10}(a),(b), $n=l_s/(2T|v_g^F(k^*)|)$ are marked by
vertical dotted lines at $l_s=200$ in the main panels for the $k^*$
where $D_g^F(k)$ diverges, and we indeed see that the local peaks of
the mutual information are in their neighborhood.

\begin{figure}
\centering
        \subfigure[]{%
            \label{fig:first}
            \includegraphics[width=0.50\textwidth]{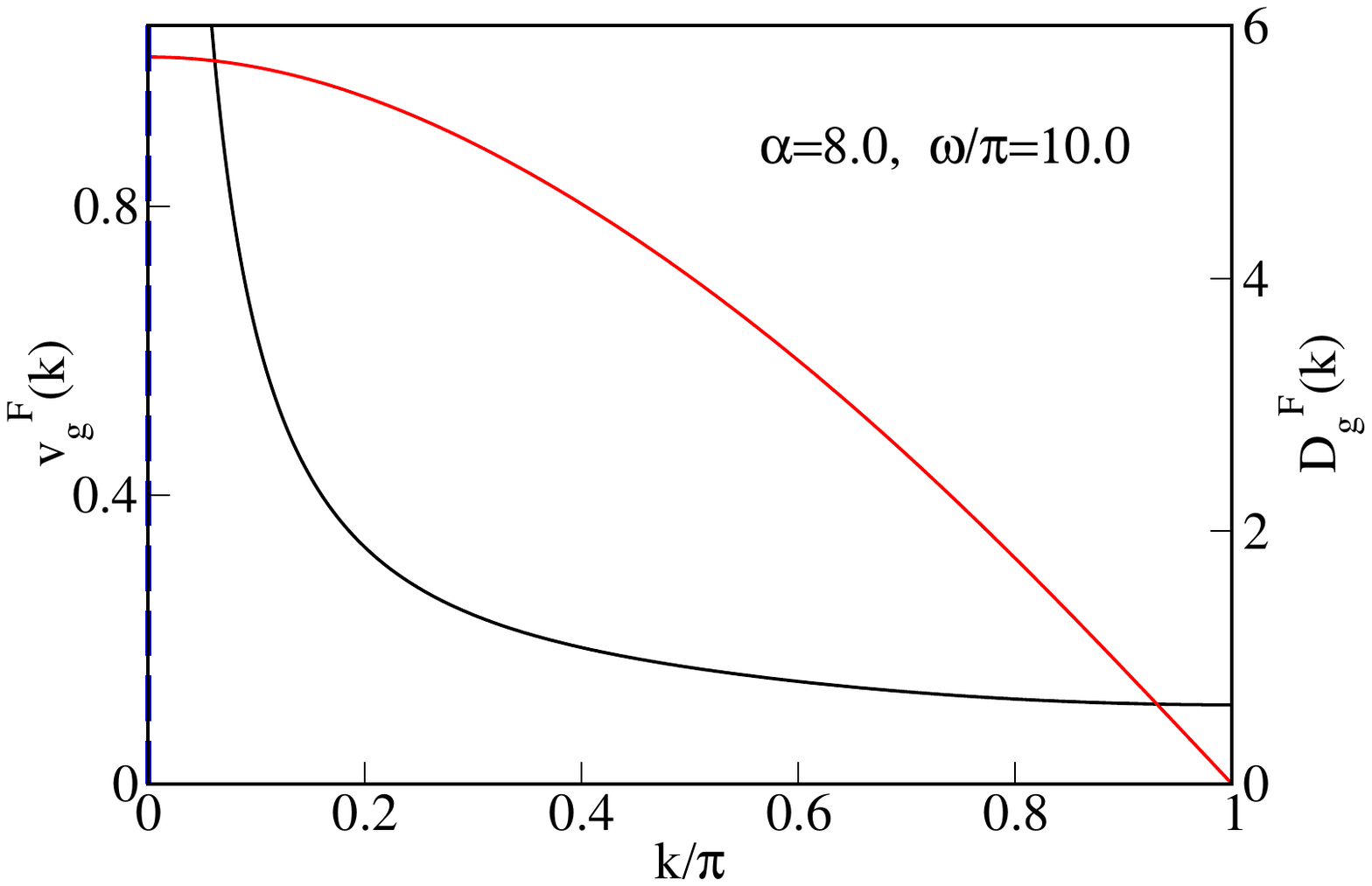}
        } \\%
        \subfigure[]{%
           \label{fig:second}
           \includegraphics[width=0.50\textwidth]{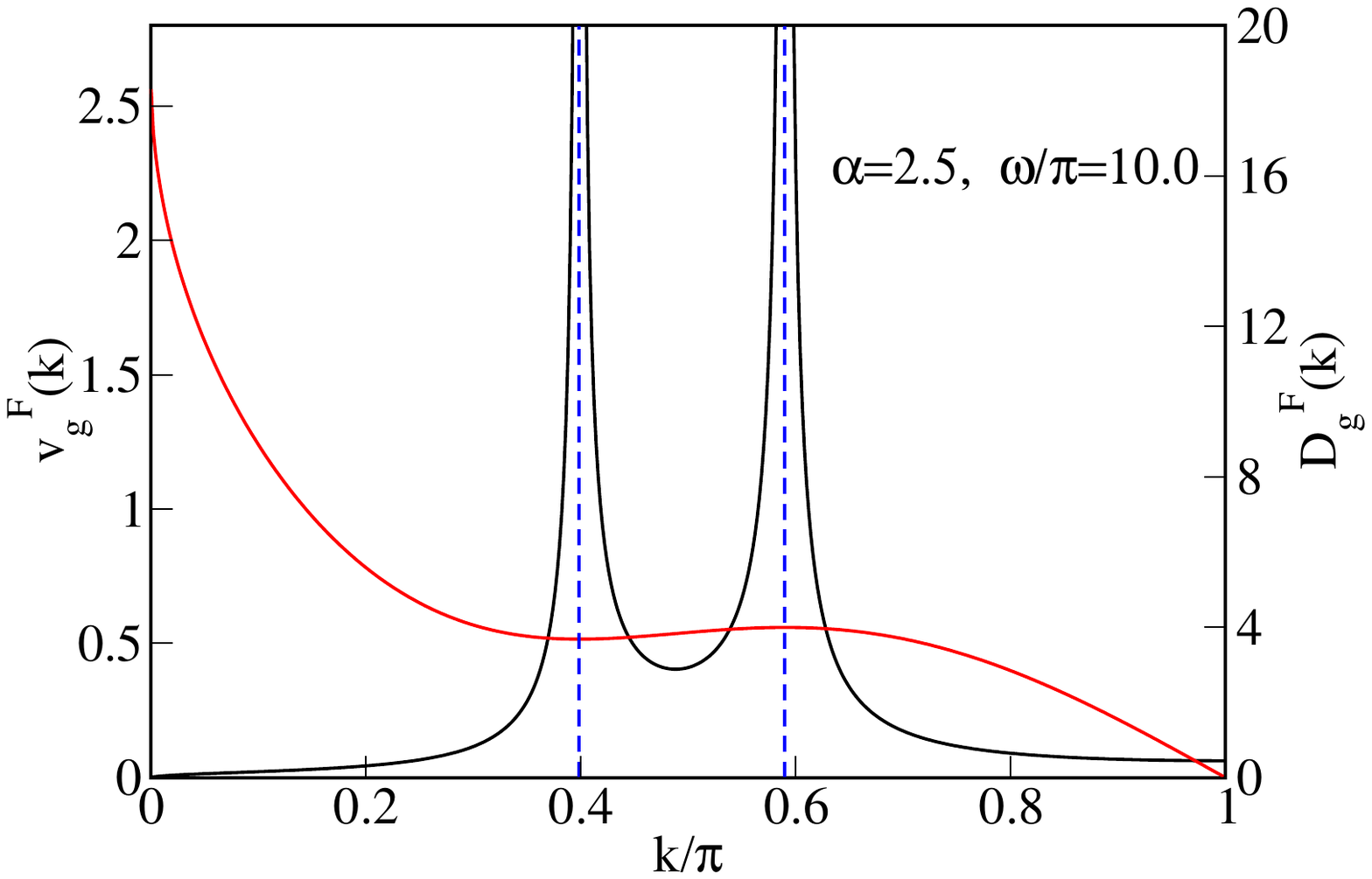}
        } 

    \caption{%
The behavior of $v_g^F(k)$ (shown in red) and $D_g^F(k)$ (shown in black)
at high frequency ($\omega/\pi=10.0$) for
(a) $\alpha=8.0$ and (b) $\alpha=2.5$. The other drive parameters are $g_i=2$ and
$g_f=0$. The locations of the divergences of $D_g^F(k)$ are shown as dotted
(blue) lines.
     }%
   \label{fig14}
\end{figure}

Let us first consider the case when $\omega \gg 1$. For $\alpha=8.0$
with $g_i=2$ and $g_f=0$, $D_g^F(k)$ has a single divergence at
$k=0$ for $\omega/\pi=10.0$ which is also the momentum $k$ at which
the Floquet group velocity $v_g^F(k)$ attains its maximum magnitude
(Fig.~\ref{fig14}(a)). This explains the simple behavior of
$\mathcal{I}_n(l,l_s)$ as shown in Fig.~\ref{fig9}(a) where there is
a single sharp mutual information front soon after it turns non-zero
as a function of $n$. Lowering the value of $\alpha$ to $2.5$
(keeping the other parameters the same as before) already leads to
an interesting difference. $D_g^F(k)$ now has two divergences, both
at non-zero values of $k$, but the maximum of $v_g^F(k)$ is still at
$k=0$ (Fig.~\ref{fig14}(b)), where $D_g^F(k)$ goes to zero. This
explains the marked difference of $\mathcal{I}_n(l,l_s)$ for
$\alpha=2.5$ (Fig.~\ref{fig9}(b)) compared to $\alpha \gg 1$. The
mutual information is suppressed in the neighborhood of $n=n_c$
(Eq.~\ref{nc}) because of the low density of quasiparticles that
have velocities close to $(v_g^F)^{\mathrm{max}}$. Instead, the peak
feature in the mutual information in Fig.~\ref{fig9}(b) is from the
contribution of the quasiparticles in the neighborhood of $k^*$ for
which $D_g^F(k) \rightarrow \infty$ here (Fig.~\ref{fig14}(b)) and
therefore, has a velocity $v_g^F(k^*)$, which is completely
different from $(v_g^F)^{\mathrm{max}}$.

\begin{figure}
\centering
        \subfigure[]{%
            \label{fig:third}
            \includegraphics[width=0.50\textwidth]{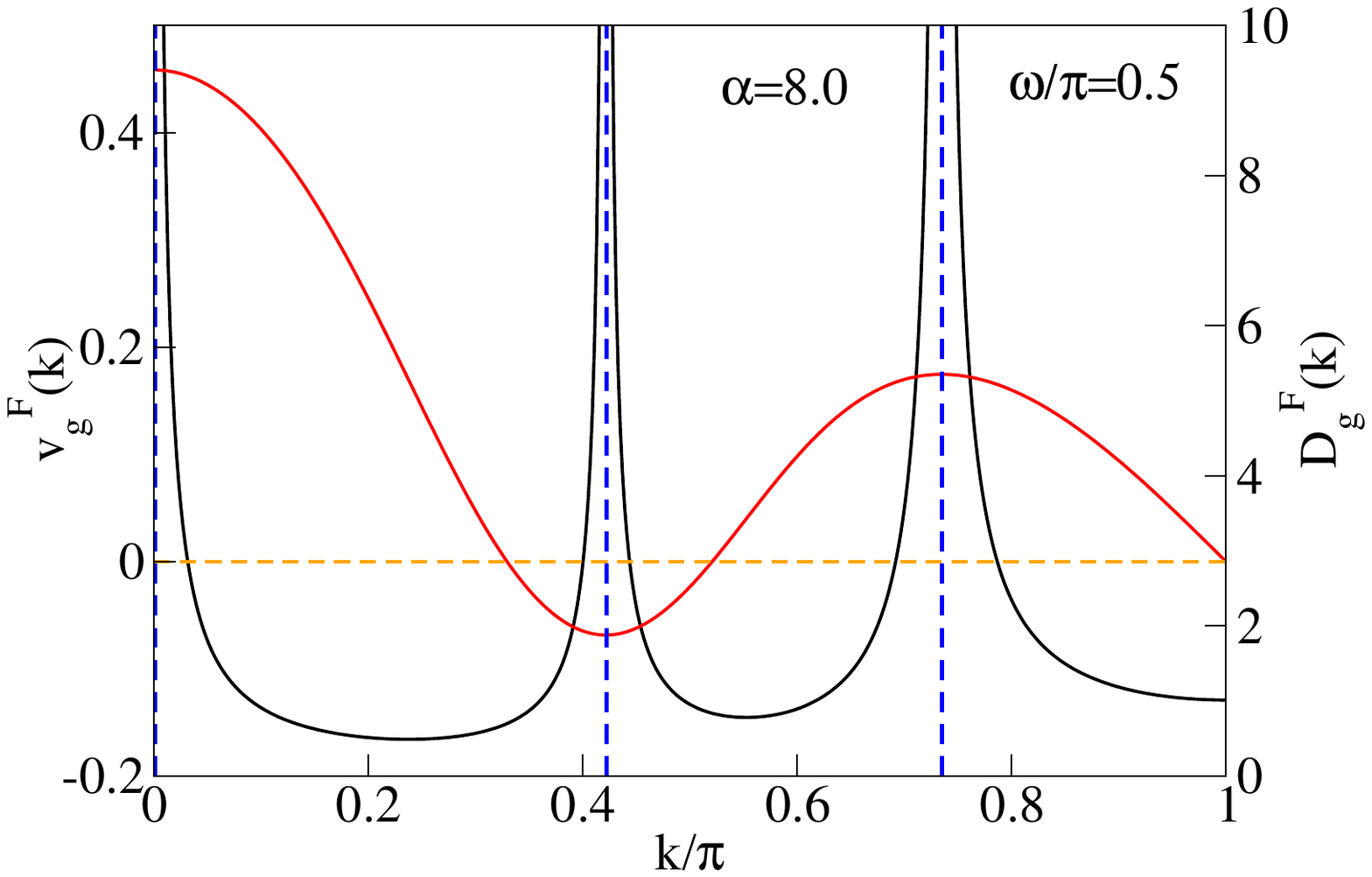}
        } \\%
        \subfigure[]{%
            \label{fig:fourth}
            \includegraphics[width=0.50\textwidth]{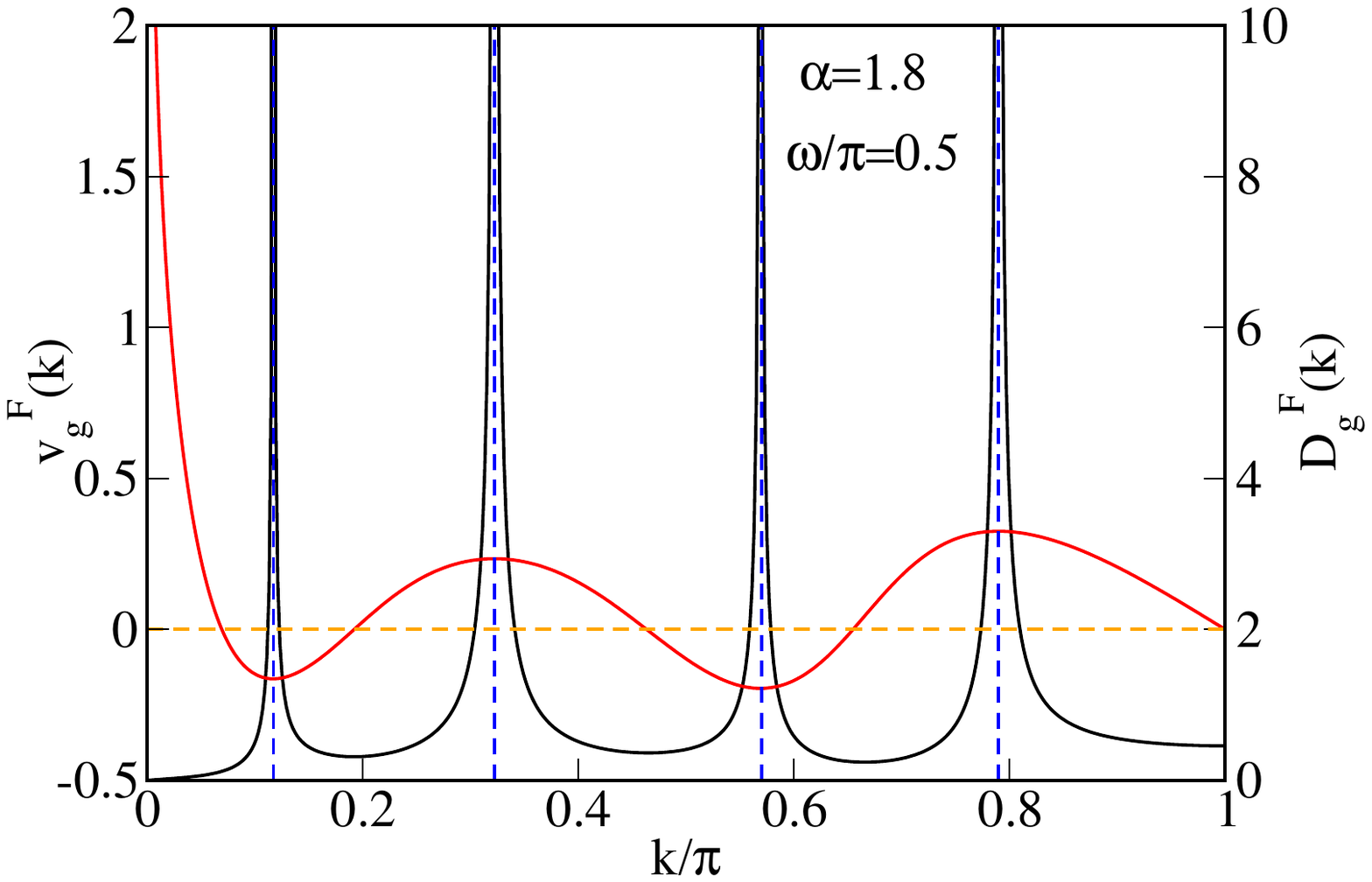}
        }%
    \caption{%
The behavior of $v_g^F(k)$ (shown in red) and $D_g^F(k)$ (shown in black)
at a drive frequency of $\omega/\pi=0.5$ for
(a) $\alpha=8.0$ and (b) $\alpha=1.8$.
The other drive parameters are $g_i=2$ and
$g_f=0$. The locations of the divergences of $D_g^F(k)$ are shown as dotted
(blue) lines.
     }%
   \label{fig15}
\end{figure}

Importantly, $D_g^F(k)$ is strongly sensitive to the driving
frequency $\omega$. When $\omega \rightarrow 0$, $v_g^F(k)$ crosses
zero a large number of times ($\sim 1/\omega$ or larger) in the BZ
as can be seen from Fig.~\ref{fig3}. Since $v_g^F(k)$ is continuous
in $k$, this implies that the number of divergences in $D_g^F(k)$
also scales in the same manner at small $\omega$, which is
qualitatively different from the behavior of $D_g^F(k)$ at large
$\omega$. We show the behavior of $v_g^F(k)$ and $D_g^F(k)$ at a
driving frequency of $\omega/\pi=0.5$ for $\alpha=8.0$
(Fig.~\ref{fig15}(a)) and for $\alpha=1.8$ (Fig.~\ref{fig15}(b))
where the other parameters are $g_i=2$ and $g_f=0$. The multiple
light cones in Fig.~\ref{fig10}(a) for $\alpha=8.0$ can now be seen
as the direct consequence of extra divergences in $D_g^F(k)$ apart
from at $k=0$ when $\omega$ is decreased. The first light cone front
as a function of $n$ arises from the quasiparticles around $k=0$
where $v_g^F(k)$ attains its maximum. However, the other two
pronounced light cone fronts in $\mathcal{I}_n(l,l_s)$ (as shown in
Fig.~\ref{fig10}(a)) are because of the quasiparticles around
$k^*_1$ and $k^*_2$, that propagate with the corresponding
$v_g^F(k)$ (Fig.~\ref{fig15}(a)), which are the other momenta where
$D_g^F(k)$ diverges. Similarly, the difference in the behavior of
$\mathcal{I}_n(l,l_s)$ for $\alpha=1.8$ at the driving frequencies
of $\omega/\pi=10.0$ (Fig.~\ref {fig9}(c)) and $\omega/\pi=0.5$
(Fig.~\ref{fig10}(b)) can again be attributed to the presence of
extra divergences in $D_g^F(k)$ as the driving frequency is varied
(Fig.~\ref{fig15}(b)). Thus, extra divergences in $D_g^F(k)$ as the
frequency is reduced from $1/\omega=0$ causes the appearance of
qualitatively new features that are absent in the global quench case
(or equivalently, at high driving frequencies). We also note here
the presence of additional local extrema in the mutual information
$\mathcal{I}_n(l,l_s)$ for both large $\alpha$ (Fig.~\ref{fig10}(a))
and for small $\alpha$ (Fig.~\ref{fig10}(b)) which cannot be simply
explained by the divergences in $D_g^F(k)$ when the driving
frequency is small. It will be useful to understand this full
structure in detail in future work.
\begin{figure}
\centering
        \subfigure[]{%
            \label{fig:first}
            \includegraphics[width=0.50\textwidth]{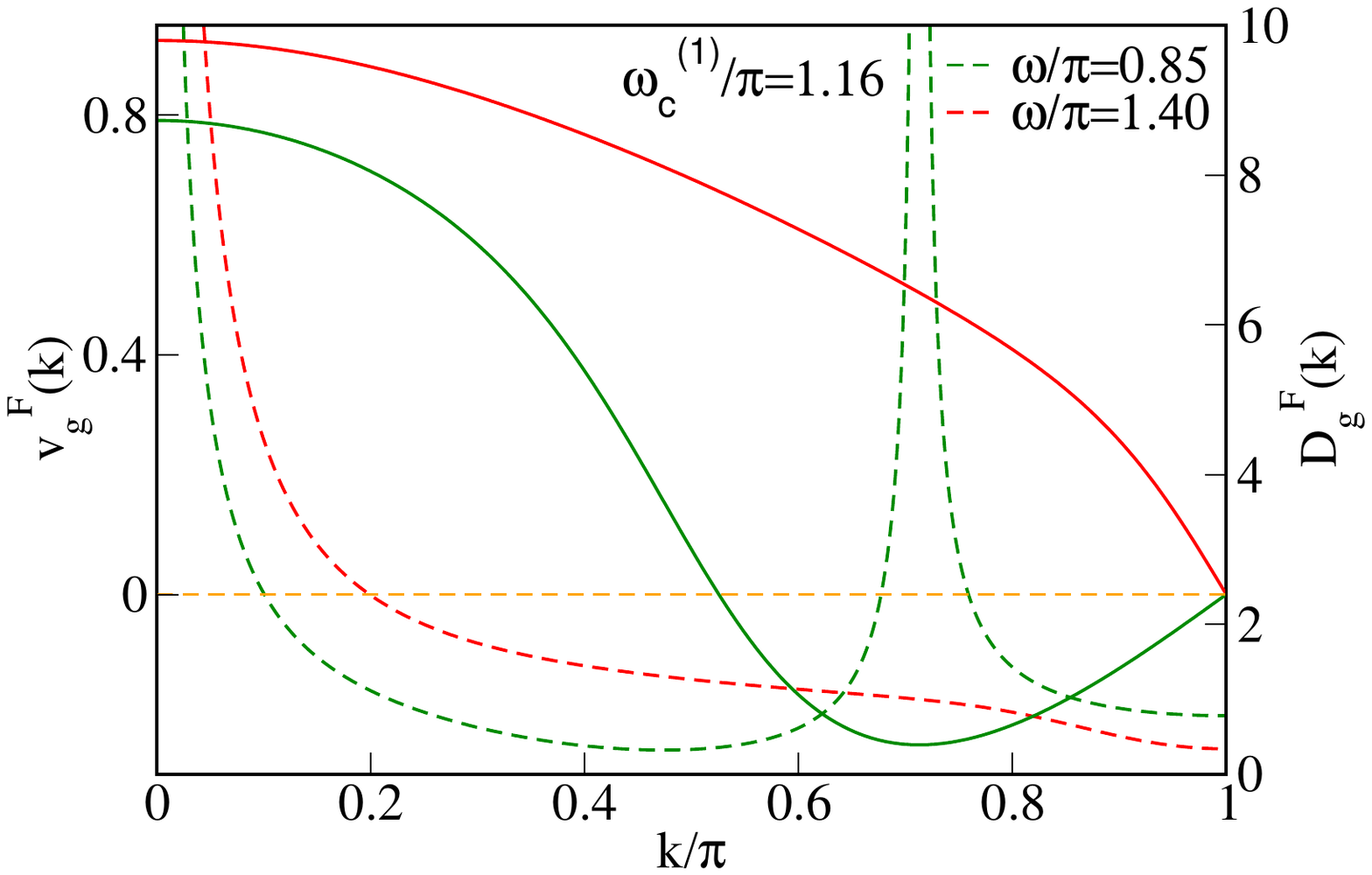}
        } \\%
        \subfigure[]{%
           \label{fig:second}
           \includegraphics[width=0.50\textwidth]{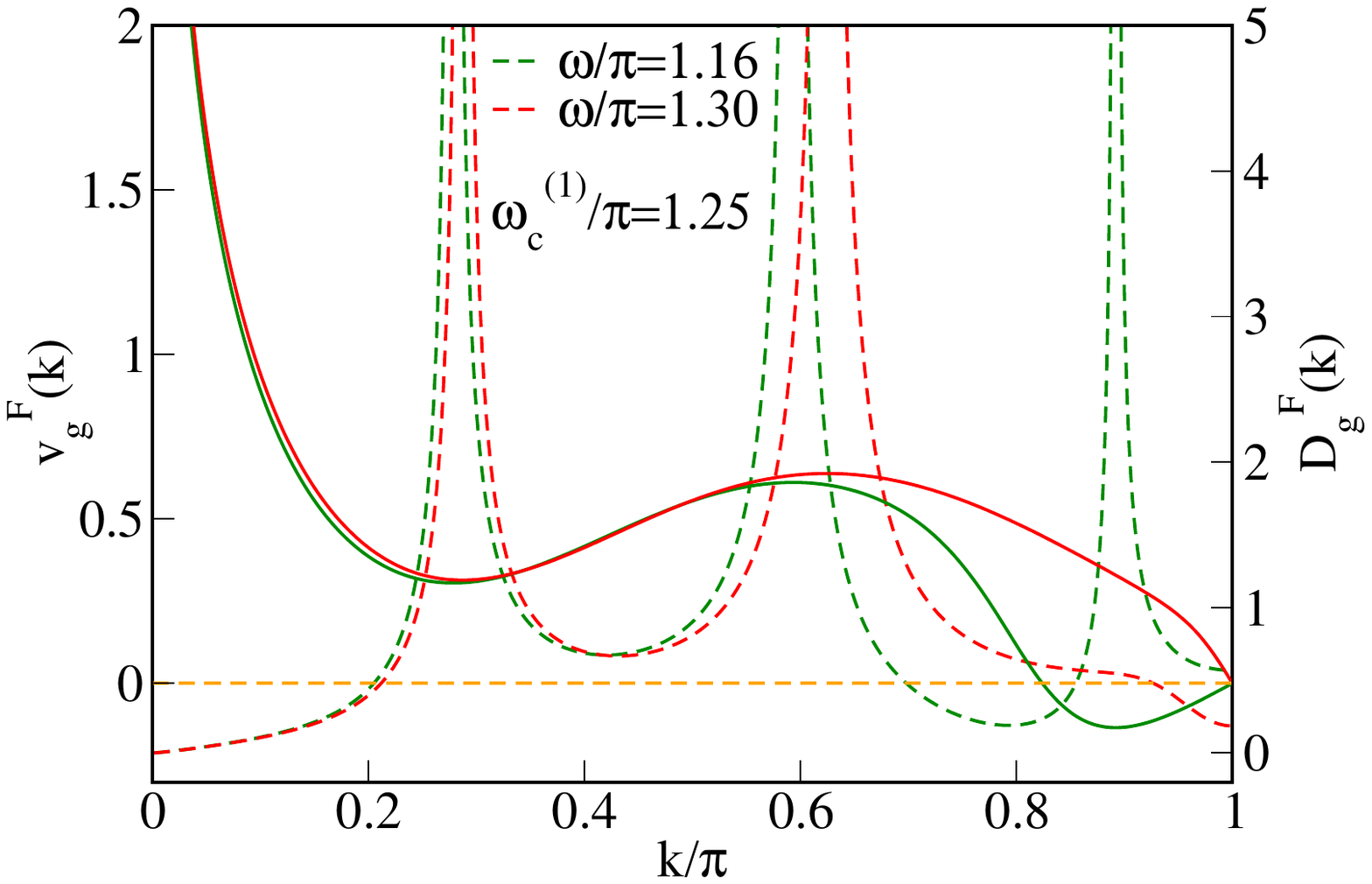}
        } 

    \caption{%
The behavior of $v_g^F(k)$ (filled lines) and $D_g^F(k)$ (dotted lines)
shown for (a) $\alpha=8.0$ and (b) $\alpha=1.8$ where the drive parameters
are $g_i=2$ and $g_f=0$. Above $\omega_c^{(1)}$, no new divergences are
produced in $D_g^F(k)$ compared to the global quench case ($\omega
\rightarrow \infty$), whereas below $\omega_c^{(1)}$, an additional
divergence is generated in $D_g^F(k)$ in both the cases.
     }%
   \label{fig16}
\end{figure}

When $\alpha > \alpha_c$, we see that {\it no new divergence}
develops in $D_g^F(k)$ compared to the global quench case ($\omega
\rightarrow \infty$) for any $\omega \in (\omega_c^{(1)},\infty)$
and an extra divergence is immediately generated for $\omega
\rightarrow \omega_c^{(1)}$ from below irrespective of whether
$\alpha>2$ (Fig.~\ref{fig16}(a)) or $\alpha<2$
(Fig.~\ref{fig16}(b)). The number of zeroes of both the functions,
$v_g^F(k)$ and $dv_g^F(k)/dk$ in $k \in [0,\pi]$ stay unchanged when
$\omega$ is above $\omega_c^{(1)}$. Just below $\omega_c^{(1)}$, an
additional zero in $v_g^F(k)$ first enters from one of the BZ edges
which causes $v_g^F(k)$ to change sign in that $k$ neighborhood
(either around $k=0$ or $k=\pi$ depending on where the new zero
enters from). Moreover, it also causes $v_g^F(k)$ to develop an
additional extremum between its new zero and the zero at the BZ
edge. Hence, an additional divergence is immediately produced in
$D_g^F(k)$ when $\omega$ goes infinitesimally below
$\omega_c^{(1)}$. As $\omega$ is lowered further, additional
divergences get generated in $D_g^F(k)$ at other specific values of
$\omega$ (because the quantity is integer-valued) since ultimately
the number of these divergences diverges as $\omega \rightarrow 0$
as discussed before. Thus, the mutual information propagation can
attain a qualitatively different profile in space-time due to
additional divergences in the function $D_g^F(k)$ when $\omega$ is
outside the range $(\omega_c^{(1)}, \infty)$, whereas inside this
frequency range, there is no qualitative distinction compared to the
case of a global quantum quench. This establishes the
presence of a sudden change in mutual information ${\mathcal
I}_n(l,l_s)$ as a function of $\omega$ at the largest dynamical transition
frequency $\omega_c^{(1)}$ for $\alpha > \alpha_c$.


\section{Conclusions and outlook}
\label{conclude}

In this work, we have analyzed a driven generalized Kitaev chain
where the degrees of freedom are spinless fermions with a nearest
neighbor hopping, an onsite chemical potential and long-ranged
p-wave pairing terms whose decay in space is characterized by an
exponent $\alpha$ (described by Eq.~\ref{Hamiltonian}).  The system
is driven by a purely unitary dynamics generated from the
time-dependence of the chemical potential $(g(t))$ that is
periodically varied in time with a frequency $\omega$. Short-ranged
integrable models with free fermion representations are known to
asymptotically synchronize with the driving frequency such that when
local (in space) properties are observed stroboscopically in time
(i.e., when the time intervals are separated by an integer multiple
of the time period ($T$) of the drive such that $t=nT$), the late
time properties reach a steady state that can be described by a
periodic generalized Gibbs ensemble which has a volume law scaling
of entanglement instead of the well-known area law scaling for
ground states and unentangled pure states. The motivation for this
work is two- fold: (a) whether and how such a long-ranged system
reach its steady state (locally) as a function of time when driven
periodically in time and (b) how does the entanglement propagate in
space and time when the system is started from an initial
unentangled pure state (the vacuum of fermions in this study)?

Regarding the former point, we show that the local properties of
such a long-ranged integrable system always reaches an asymptotic
steady state irrespective of the value of $\alpha$ and the drive
frequency $\omega$ in the thermodynamic limit. We address how the
local properties relax to their final values as a function of the
stroboscopic time $nT$ by defining an appropriate distance measure,
$\mathcal{D}_n(l) (\in [0,1])$, which is zero iff all non-trivial
correlation functions that can be defined by using any subset of $l$
adjacent sites in the system coincide with their corresponding
values in the final steady state. We show that there are only two
possible dynamical phases when the drive frequency is varied for any
value of $\alpha$ which are characterized by either a
$\mathcal{D}_n(l) \sim (\omega/n)^{-3/2}$ or a $\mathcal{D}_n(l)
\sim (\omega/n)^{-1/2}$ behavior when $n \gg 1$ for any finite $l$
in the thermodynamic limit. We show that there exists a critical
range $\alpha_c$ that only depends on the time-averaged value of
$g(t)$ over one full drive cycle, denoted by $g_{avg}$, such that
above $\alpha_c(g_{avg})$, $\mathcal{D}_n(l) \sim (\omega/n)^{3/2}$
[$\mathcal{D}_n(l) \sim (\omega/n)^{1/2}$] as $\omega \rightarrow
\infty [0]$ whereas below $\alpha_c(g_{avg})$, $\mathcal{D}_n(l)
\sim (\omega/n)^{1/2}$ both for high and low frequency driving.
Since the problem maps on to a global quantum quench with the
post-quenched Hamiltonian equal to the time-averaged one (over one
full period of the drive) when $\omega \gg 1$, this implies that
there is a dynamical phase transition at $\alpha_c$ (keeping other
parameters fixed) with a global quench protocol. We also map out the
rich phase diagram for these dynamical phases as a function of the
drive frequency and amplitude for different values of $\alpha$ and
point out the distinctions between short-ranged ($\alpha \gg 1$) and
long-ranged ($\alpha \sim 1$) pairing terms.

Regarding the latter point, we study the mutual information
${\mathcal I}_n(l,l_s)$ which is a reliable measure of entanglement
generation as a function of $n$, $\omega$ and $\alpha$. Our study
finds qualitatively different features in ${\mathcal I}_n(l,l_s)$ as
a function of $\omega$ and $\alpha$ which can be quantitatively
understood from the properties of the Floquet group velocity
$v_g^F(k)$ and the corresponding density of states $D_g^F(k)$. We
find that for $\alpha > 2 > \alpha_c$, where at least one
dynamical transition exists at $\omega=\omega_c^{(1)}$, ${\mathcal
I}_n(l,l_s)$ exhibits a single light-cone like feature analogous to
the one obtained for quantum quenches
\cite{HaukeT2013,RegemortelSW2016,BuyskikhFSED2016} for $\omega >
\omega_c^{(1)}$. In contrast, for $\omega < \omega_c^{(1)}$, it
shows multiple light-cone like features which can be shown to be the
consequence of appearance of new zeroes in $v_g^F(k)$. The first of
such additional zeroes appear at the dynamic transition with the
highest frequency ($\omega=\omega_c^{(1)}$); the behavior of
$I_n(l,l_s)$ as a function of $n$ changes suddenly at this point
relating the dynamic transition to the behavior of ${\mathcal
I}_n(l,l_s)$. We also find that the behavior of ${\mathcal
I}_n(l,l_s)$ for $\alpha \le 2$ is fundamentally different from its
counterpart for $\alpha > 2$ at least in two major ways. First,
${\mathcal I}_n(l,l_s)$ do not exhibit a light cone structure for
any $\omega$ and second the propagation of entanglement between two
subsystems is instantaneous for $\alpha \le 2$ making $I_n(l,l_s)$
finite for any $n>0$ in contrast to its counterpart for $\alpha
> 2$ which is finite for $n>n_c$ (Eq.\ \ref{nc}). These differences
may be understood from the fact that for $\alpha \le 2$, $v_g^F(k)$
diverges at $k =0$; thus Eq.\ \ref{nc} has a solution for any $n>0$
which ensure instant propagation of entanglement. In contrast, for
$\alpha > 2$, $v_g^F(k)$ and hence $n_c$ is finite for all $k$,
leading to single or multiple light cone like features along with
finite entanglement propagation time. We note that
the fact that $n_c$ is zero for all $\alpha <2$ indicates that the
spread of mutual information can not clearly distinguish between
quasi long-range ($1 <\alpha <2$) and long range ($\alpha <1$) interaction
regimes \cite{HaukeT2013, RegemortelSW2016, BuyskikhFSED2016} in the
sense that it propagates instantaneously for any $\alpha <2$. Our
work therefore points out that the spread of entanglement in a
closed quantum system depends on both the drive frequency and the
long/short-range nature of its Hamiltonian.

To conclude, we have studied a periodically driven Kitaev chain
whose pair-potential decays in space with an exponent $\alpha$. For
$\alpha >\alpha_c$, we have found the existence of at least one
dynamic transition in this model separating two dynamical phases in
which all correlator of the system decay to their steady state
values as $(\omega/n)^{3/2}$ [$(\omega/n)^{1/2}$] for high(low)
frequencies. For $\alpha < \alpha_c$, no such transition exists and
all correlator exhibit $n^{-1/2}$ decay at all frequencies (except
for fine-tuned regions); this allows for a change in the phase of
the driven system at high frequencies by tuning $\alpha$ through
$\alpha_c$. We have also shown that the behavior of the entanglement
entropy exhibits at sudden change at the dynamic transition; at high
frequencies, the space-time behavior of the mutual information
exhibits a single light cone when $\alpha
> 2$ while at low frequencies, multiple light cones
exist. This change can be understood from an analysis of the Floquet
Hamiltonian of the system. For $\alpha_c < \alpha < 2$, even though
the entanglement propagation is instantaneous and no light cone like
features exist at any $\omega$, the behavior of the mutual
information again shows no new features when $\omega \in
(\omega_c^{(1)} , \infty)$ while qualitatively new features appear
when $\omega < \omega_c^{(1)}$. Finally, our work suggests that it
will be interesting to explore the presence of such dynamical phases
in Bethe-integrable systems \cite{bethepapers} and in the
pre-thermal regime of non-integrable models \cite{prepapers}, which
are close to integrable points, and to understand the dynamics of
entanglement spreading in aperiodically driven (both random and
quasiperiodic) integrable systems.~\cite{NandySS2017}

{\it Acknowledgements:} The work of A.S. is partly supported
through the Partner Group program between the Indian Association for the
Cultivation of Science (Kolkata) and the Max Planck Institute for the
Physics of Complex Systems (Dresden). The authors thank T.~Kuwahara 
for useful discussions.

\end{document}